\documentclass[intlimits,twoside,a4paper]{article}

\usepackage[cp1251]{inputenc}
\usepackage{graphicx}

\usepackage{cmpj3}

\issue{2020}{23}{2}{23604}
\doinumber{10.5488/CMP.23.23604}

\title[Water in pores with molecular brushes]
{Towards the description of water adsorption in slit-like nanochannels with grafted molecular brushes.
Density functional theory
}

\author[V.M. Trejos, M. Aguilar, S. Soko{\l}owski, O. Pizio]{V.M. Trejos\refaddr{label1}, M. Aguilar\refaddr{label2},  S. Soko{\l}owski\refaddr{label3}, O. Pizio\refaddr{label4}\thanks{E-mail: oapizio@gmail.com}}
\addresses{
\addr{label1} 
Instituto de Ciencias B\'asicas e Ingenier\'ia, Universidad Aut\'onoma
del Estado de Hidalgo, Carretera Pachuca-Tulancingo Km. 4.5, Col. Carboneras, C.P. 42184,
Mineral de la Reforma, Hidalgo, M\'exico
\addr{label2}
Instituto de Qu\'{i}mica, Universidad Nacional Aut\'{o}noma de M\'{e}xico,
Circuito Exterior, 04510, Cd. de M\'{e}xico, M\'{e}xico
\addr{label3} 
Department for the Modelling of Physico-Chemical Processes,
Maria Curie-Sk{\l}odowska University, \\ Lublin 20-031, Poland
\addr{label4}
Instituto de Qu\'{i}mica, Universidad Nacional Aut\'{o}noma de M\'{e}xico,
Circuito Exterior, 04510, Cd. de M\'{e}xico, M\'{e}xico }

\date{Received January 19, 2020, in final form March 10, 2020}

\begin{document}
\maketitle

\begin{abstract}
We have explored a model for adsorption of water into slit-like nanochannels with
two walls chemically modified by grafted polymer layers forming brushes.
A version of density functional method is used as theoretical tools.
The water-like fluid  model adopted from the work of Clark et al.
[Mol. Phys., 2006, {\textbf 104}, 3561] adequately reproduces the bulk vapour-liquid coexistence envelope. The polymer layer consists of chain molecules in the framework of pearl-necklace model.
Each chain molecule is chemically bonded to the pore walls by a single terminating segment.
Our principal focus is in the study of the dependence of polymer 
layer height on grafting density and in the microscopic structure of the interface
between adsorbed fluid and brushes. Thermal response of these properties upon
adsorption is investigated in detail. The results are of importance to understand shrinking 
and swelling of the molecular brushes in the nanochannels.


\end{abstract}

\section{Introduction}

This manuscript has been prepared as a tribute to Prof. Ihor Mryglod, distinguished Ukrainian
scientist in the field of statistical physics, on behalf of his 60th birthday. Dr. Mryglod has made
several important contributions along different lines of research within
the theory of liquids. Some of his works related to our scientific interests 
are  focused on the comprehension of the equilibrium and out-of-equilibrium behaviour of water 
and other liquids as well as on the relation between the collective variables
method and the density functional approaches in the theory of inhomogeneous fluids,
see e.g.,~\cite{ihor1,ihor2,ihor3}.

Synthesis and design of smart interfaces is one of the important areas of applied 
research and a challenging subject for theoretical investigations.
Grafted polymer brushes represent an essential ingredient of this type of materials.
Design of adaptive surfaces requires the use of malleable polymers whose properties change 
in response to environmental stimuli such as temperature,  electric field, pH or 
light~\cite{jonas1,jonas2,conrad,li,minko1,minko2,minko3,slavko1,constable1,constable2}. 
Profound comprehension of surface/interface properties is a 
prerequisite of developing rational synthetic strategies and 
new structures with functionalities for novel applications~\cite{conrad,li,constable1,constable2}. 

Conformational behaviour of grafted polymer layers is
determined by the length of chain molecules, grafting density, and the curvature of
the supporting substrate. Usually, three regimes are distinguished in 
the brush formation, namely, mushroom type structure,  crossover regime and 
 highly stretched regime.
Specifically, at low grafting density and with poor solvent, the polymer chains
prefer to be collapsed. With an increasing grafting density, a semi-dilute polymer brush
regime is realized, the configurational entropy of chains becomes reduced compared to
mushroom-like structures. At  high grafting densities,
polymer chains become  stretched due to an increased mutual confinement 
that creates a rather concentrated polymer brush layer~\cite{minko3,ethier}.

From  theoretical perspective, computer simulation methods, see e.g.,
\cite{malfreyt,dimitrov1,dimitrov2,dimitrov3,goicochea,ii10,ii11,slavko2,slavko3} and \cite{slavko4} for
a quite recent review, and entirely theoretical approaches
have been applied to  describe systems involving tethered brushes. The latter include
the  self-consistent field approaches \cite{milner,int6,int7,int8,int9}, 
and density functional (DF) theories~\cite{chapter}. 
The DF approaches represent versatile tools to describe an ample variety of
brush-fluid models and have been applied in the studies of
lipids, co-polymers, grafted polymers, polymer/colloid, and polymer/nanoparticle 
systems~\cite{Chapman2007,Chapman2008,Chapman2011,Chapman2012,Chapman2013}.
In particular, Yu and Wu \cite{c40,c41} developed a successful 
version of the DF theory to describe systems of nonuniform polymers.   

In this work, we present a continuation of the project,
see \cite{trejos-x1,trejos-x2,trejos-x3,trejos-x4,trejos-x5}, 
focused on the study of the behaviour of water in complex porous media.
At the first stage, in \cite{trejos-x1}, we  studied thermodynamic properties of 
a set of water-like models designed in \cite{clark} by applying the method similar to~\cite{gil}. 
The principal idea behind the modelling of~\cite{clark} is to reproduce the liquid-vapour (LV) coexistence 
of water using square-well attraction and site-site chemical association
without resorting to electrostatic inter-particle interactions.
We incorporated this model into a DF approach to describe the behaviour of water-like models 
in slit-like pores~\cite{trejos-x2}. 
Our theoretical findings appeared to be in agreement with computer simulation results of much more 
sophisticated water models in pores~\cite{Ulberg1994,Ulberg1995,Cummings2011}. 
The most recent part of our studies along this line of research is focused on
the effects of chemical modification of slit-like pore walls by 
grafted polymer layers~\cite{trejos-x3,trejos-x4,trejos-x5}.
Specifically, we studied the adsorption of
water in slit-like pores with grafted polymer layers on the walls resorting to 
a version of the DF theory. Shrinking or swelling of molecular brushes 
induced by  phase transitions in confined 
water~\cite{trejos-x5} was explored.

In the present contribution we continue the investigation of the model
from \cite{trejos-x3,trejos-x4,trejos-x5} concerned with the properties of water
in pores with chemically modified walls. However, the principal purpose
of this work is to describe the thermal response of grafted polymer layers
to the conditions of adsorption. In contrast to~\cite{trejos-x3,trejos-x4,trejos-x5},
we assume that there is an attractive interaction between segments of grafted
chain molecules. Therefore, the brushes are thermal on their own, not
only due to the action of the adsorbed fluid. This modelling is closer
to the experimental setup~\cite{conrad} and permits to obtain additional 
insights into the behaviour of this class of adsorption systems.

The manuscript is organized as follows. In sections \ref{model} and \ref{theory},
the setup is described as briefly as possible, because a detailed description
of the model and theory has been given elsewhere~\cite{trejos-x5}. 
In addition, a few essential equations are given in the supplementary material 
for the sake of convenience of the reader.
In section \ref{results}, the original results are discussed. Next, the conclusions
and summary are given.
\section{Model}\label{model} 

We consider the fluid model as one-component fluid of associating molecules. 
Each fluid molecule has four associative sites designated by A, B, C and D,
inscribed into a spherical core~\cite{nezbeda1,jackson1,jackson2}.
The set of all the sites is denoted by $\Gamma$. 
The pair inter-molecular potential between molecules 1 and 2 depends
on the center-to-center distance and orientations,
\begin{equation}
 u(12) = u_\text{ff}(r_{12}) + \sum_{\alpha \in \Gamma} \sum_{\beta \in \Gamma} 
u_{\alpha\beta}(\mathbf{r}_{\alpha\beta}),
\end{equation}
where
$\mathbf{r}_{\alpha\beta}=\mathbf{r}_{12}+\mathbf{d}_{\alpha}(\omega_1)-\mathbf{d}_{\beta}(\omega_2)$
is the vector that connects site $\alpha$ on molecule 1 with site $\beta$ on molecule~2,
 $r_{12}=|\mathbf{r}_{12}|$ is the distance between the centers of molecules 1 and 2,
$\mathbf{\omega}_i$ is the orientation of the molecules~$i$, $\mathbf{d}_{\alpha}$
is the vector from the molecular center to site $\alpha$, see also figure~1 of \cite{jackson1}.
Each of the off-center attraction sites is located at a
distance $d_\text{s}$ from the particles' center, $d_\text{s} = |\mathbf{d}_{\alpha}|$
($\alpha = \text{A, B, C, D}$).
In the model in question, only the site-site association  AC,  BC, AD, and BD is allowed,
all association energies are  equal. Specifically, the interaction between sites
is given as
\begin{equation}
\label{eq:asw}
u_{\alpha\beta}(\mathbf{r}_{\alpha\beta})=
\left\{
\begin{aligned}
&-\varepsilon_\text{as}\,,~~~~\,\text{if}~~    0< |\mathbf{r}_{\alpha\beta}| \leqslant r_\text{c} \,,\\
&0,~~~~~~~~~~~~\text{if}~~     |\mathbf{r}_{\alpha\beta}|   > r_\text{c} \,,\\
\end{aligned}
\right. 
\end{equation}
where $\varepsilon_\text{as}$ is the depth of the association energy well and
$r_\text{c}$ is the cut-off range of the associative interaction. Note that the same
model was also used in several works.

The non-associative part of the pair potential, $u_\text{ff}(r)$, is given as,
\begin{equation}
u_\text{ff}(r) = u_\text{hs,ff}(r) +  u_\text{att,ff}(r),
 \label{eq:sw}
\end{equation}
where $u_\text{hs,ff}(r)$ and $u_\text{att,ff}(r)$ are the
hard-sphere (hs) and attractive (att) pair interaction potential, respectively.
The hs term is,
\begin{equation}
u_\text{hs,ff}(r) =
\left\{
\begin{aligned}
&\infty, ~~~~~~~~\,\text{if}~~r < \sigma,\\
&0,~~~~~~~~~~\text{if}~~r \geqslant \sigma,
\end{aligned}
\right.
\end{equation}
where $\sigma$ is the hs diameter.
The attractive interaction is described by the square-well (SW) potential,
\begin{equation}\label{uSW}
 u_\text{att,ff}(r)=
\left\{
\begin{aligned}
&0,~~~~~~~~~\,\text{if}~~ r < \sigma,\\
&-\varepsilon,~~~~~~\text{if}~~ \sigma \leqslant r < \lambda_\text{ff} \sigma,  \\ 
&0,~~~~~~~~~~\text{if}~~r \geqslant \lambda_\text{ff} \sigma,
\end{aligned}
\right.
\end{equation}
where $\varepsilon$ and $\lambda_\text{ff}$ are the depth and the range of the potential, respectively.

The pore walls are located at $z=-H/2$ and $z=H/2$.
The external potential, $v(z)$, exerted  on a fluid particle inside the pore
by the ``bare'' (non-modified) walls is,
\begin{equation}\label{Eqv_z}
v_\text{f}(z) = v_\text{fw}(H/2+z)  + v_\text{fw}(H/2-z) {\ \ \ \rm for \ \ \ } -H/2\leqslant z\leqslant H/2.
\end{equation}
The function $v_\text{fw}(z)$ is given by the Steele's 10-4-3 gas-solid potential \cite{steele,steele2},
\begin{equation}
 v_\text{fw}(z) =\varepsilon_\text{fw}
 \left[ \frac{2}{5} \left( \frac{ \sigma_\text{fw}}{z}\right)^{10} 
 - \left( \frac{ \sigma_\text{fw}}{z}\right)^{4} \right. 
\left. - \frac{\sigma_\text{fw}^4 }
{3 \Delta (z+0.61 \Delta)^3 }
 \right],
\end{equation}
where $\varepsilon_\text{fw}$, $\sigma_\text{fw}$ are
the energy and the size parameters describing fluid-wall interaction ($\Delta$ is the solid inter-layer 
spacing). 

In order to describe the layer of tethered chain molecules at each wall we use the approach 
presented already in \cite{caowu,c21,c22}.
The grafted layer at one wall is composed of chains of $M_1$ tangentially jointed segments,
while each chain attached to the second wall comprises $M_2$ segments. For simplicity,
we assume that the diameter of all the segments is the same and equal to $\sigma_\text{cc}$.
Moreover, we restrict ourselves to a symmetric case $M_1=M_2=M$. 

The connectivity of segments in the chain $I$ is provided by
imposing the bonding potential, $V_\text{b}^{(I)}$, see~\cite{c40,c41}
\begin{equation} \label{eq:1}
\exp \big[-V_\text{b}^{(I)}({\bf R})/kT\big]=\prod_{i=1}^{{M_I}-1}\delta (|{\bf r}_{i+1}-%
{\bf r}_{i}|-\sigma_\text{cc})/4\piup (\sigma_\text{cc})^{2},
\end{equation}
where $I=1,2$; ${\bf R}_k\equiv (\mathbf{r}_{1},\mathbf{r}_{2},\ldots ,\mathbf{r}%
_{M})$ is the vector that specifies the coordinates of all segments and  $\delta(x)$ 
the Dirac function.

The first segment of the chain $I=1$  is bonded to the wall  at $z=-H/2$, 
whereas the first segment of the chain $I=2$ is bonded to the wall at $z = H/2$.
It means that these segments are fixed in the $z$ plane at the positions 
$z = \sigma_\text{cc}/2 - H/2$ and $z = H/2 - \sigma_\text{cc}/2$,
respectively, or that the corresponding Boltzmann factors associated with the external potential
are as follows
\begin{equation} 
\exp\big[-v_{\text{s}1}^{(1)}(z)/kT\big]= \delta (z-(\sigma_\text{cc}/2-H/2)),
\end{equation}
and
\begin{equation} \label{eq:3}
\exp\big[-v_{\text{s}1}^{(2)}(z)/kT\big]= \delta (z-(H/2-\sigma_\text{cc}/2)).
\end{equation}
All the remaining segments of grafted chains are assumed to interact with the pore walls 
similarly to fluid species, namely,
\begin{equation}
v_{\text{s}i}^{(I)}(z)=
\left\{
\begin{array}{ll}
 \infty, & z <-H/2+\sigma_\text{cc}/2, \quad z>H/2-\sigma_\text{cc}/2, \\
v_\text{s}(z), & -H/2+\sigma_\text{cc}/2 \leqslant z\leqslant H/2-\sigma_\text{cc}/2,
\end{array}
\right. 
\label{eq:hw}
\end{equation}
where the function $v_\text{s}(z)$  is given by equation~(\ref{Eqv_z}), though with the parameters 
$\varepsilon_\text{bw}$, $\sigma_\text{bw}$, to abbreviate brush-wall interaction.
In the present contribution, we assume $v_\text{s}(z)=0$. This setup 
for segment-phobic walls makes the model simpler
and makes interpretation of the results more transparent. Possible
effects of attraction between segments and pore walls on the properties
of the systems in question will be considered elsewhere.

The inter-particle interaction between all segments as well as 
between adsorbate molecules and each segment of the grafted chain
is assumed  in the form of the 
SW potential like  it is given by equations~(\ref{eq:sw})--(\ref{uSW}) 
with size and energy parameters equal
to $\sigma_\text{cc}$, $\lambda_\text{bb}$ and $\varepsilon_\text{bb}$, 
$\sigma_\text{bf}$, $\lambda_\text{bf}$ and $\varepsilon_\text{bf}$,
respectively. The presence of attraction between segments makes the 
grafted polymer layers thermal on their own.

\section{Theory}\label{theory}

The system is studied using the version of the density functional theory (DF),
described already in detail in   \cite{caowu,c21,c22,c40,c41}. To avoid unnecessary
repetition, we recall only the basic equations. 
A few  additional equations are given in the supplementary material section.

Let us introduce the following notation. The symbols  
 $\rho^{(\text{c}I)}({\bf R})$ and  $\rho({\bf r})$ denote
the local density of the chains tethered at the wall ($I=1,2$) and of the fluid, respectively.
We also define the local densities of consecutive chains' segments,
$\rho _{\text{s}j}^{(I)}({\bf r})$, and the total segment density, $\rho _\text{s}({\bf r})$, as in~\cite{c40},
\begin{equation}  \label{eq:6}
\rho _\text{s}^{(I)}({\bf r})=\sum_{j=1}^{M}\rho _{\text{s}j}^{(I)}({\bf r}%
)=\sum_{j=1}^{M_I}\int \!\!\rd\mathbf{R}\,\delta (\mathbf{r}-{\bf r}%
_j)\rho^{(\text{c}I)}({\bf R})\;,
\end{equation}
where ${\bf R}\equiv (\mathbf{r}_{1},\mathbf{r}_{2},\ldots ,\mathbf{r}%
_{M})$ is a set of coordinates describing the positions of all segments of a given
chain molecule. 
In the model with the external potential dependent solely on the distance from the
pore walls and under the assumption of random distribution of tethered segments,
the density profiles are one-dimensional. 

Before proceeding to the essence of the theory, it is worth to make the following comments.
In the experimental setup, the polymer brushes are prepared through two techniques referred to 
as as ``grafting-to'' and ``grafting-from''.
The {\it grafting-to} method involves adsorption of pre-formed polymers to a substrate.
On the other hand,  in the {\it grafting-from}  method, the polymer chains grow from the substrate
modified by a certain initiator \cite{conrad,li}.
In our gedankenexperiment mimicking {\it grafting-from} setup,
the grafted polymer layers are present in the pore prior to water adsorption.
Thus, a combined system, slit-like pore with tethered chains and  with
adsorbed fluid, is in equilibrium with the external reservoir containing solely a fluid
at a given chemical potential or equivalently at a certain external pressure.

Consequently, in order to evaluate the density profiles we minimize the 
functional of the following thermodynamic potential~\cite{c21,c22},
\begin{equation}
{\cal Y} = F[\rho ^{(\text{c}1)}({\bf R}), \rho ^{(\text{c}2)}({\bf R}), \rho({\bf r})]+
A_\text{s}\int\!\!\rd{z}\,\rho(z)[v_\text{f}({z})-\mu],
\end{equation}
where $F[\rho ^{(\text{c}1)}({\bf R}),\rho ^{(\text{c}2)}({\bf R}), \rho ({\bf r})]$ is the Helmholtz free energy
functional and $\mu$ is the chemical potential of the fluid.
The external field, $v_\text{f}(z)$, is given by equation~(\ref{Eqv_z}) and $\mu$ is the chemical potential.
 Minimization of the thermodynamic potential $\cal Y$ is performed
under the constraint that the amount of tethered chains at each wall is constant,
\begin{equation}
\int  \rd\mathbf{r}\,\rho_{\text{s}1}^{(I)}(z)= A_\text{s} R_\text{c}^{(I)},
\label{eq:con}
\end{equation}
where $A_\text{s}$ is the surface area and $R_\text{c}^{(I)}$ is the brush density at the wall ($I=1,2$).
For simplicity, we restrict ourselves to symmetric systems with $R_\text{c}=R_\text{c}^{(1)}=R_\text{c}^{(2)}$.

The free-energy functional consists of the ideal, $F_\text{id}$, and 
the excess, $F_\text{ex}$, term. 
 The excess free energy is expressed as the sum of
contributions arising from different kinds of interactions in the system~\cite{YuWu2002,ca2,ca3}.
The definition of the free energy of the system is 
given in the supplementary material. We only mention that the 
excess contribution due to chemical association is managed by using Wertheim's theory of
association~\cite{wertheim1,wertheim1_,wertheim2,wertheim2_}. It
involves the function $\chi_\text{A}(z)$, the density profile of the fraction of molecules at 
the $z$ position  that are not bonded at the site A, which is a local analogue
of the mass action law for inhomogeneous chemically associating fluids.

The density profiles are obtained from  minimization of the 
functional ${\cal Y}$. Minimization is carried out under the 
constraint given by equation~(\ref{eq:con}).
\begin{equation}
 \frac {\delta {\cal Y}}{\delta \rho^{(\text{c}1)}({\bf R})}
 = \frac {\delta {\cal Y}}{\delta \rho^{(\text{c}2)}({\bf R})}
 =0, {\rm \ \ \ \ \ \ \ \ } \frac {\delta {\cal Y}}{\delta \rho(z)}=0.
 \label{15}
\end{equation}
Final equations for the density profiles are given in the supplementary material.

The vapour-liquid coexistence for a fluid confined in slit-like pores is found 
from the condition of equality of the thermodynamic potential ${\cal Y}$
 for two different  density profiles at a fixed temperature 
and chemical potential (cf. \cite{c21,c22}). 
The average densities of the coexisting phases of the fluid confined in a pore, $\langle\rho\rangle$,
and the average fraction of non-bonded fluid particles, $\langle\chi_\text{A}\rangle$,
are calculated from the density profiles as
\begin{equation}
 \langle\rho\rangle = \frac{1}{H} \int_{-H/2}^{H/2}  \rho(z) \rd z,  {\rm \ \ \ \ \ \ \ \ }
  \langle\chi_\text{A}\rangle=\frac{1 }{H} \int_{-H/2}^{H/2}  \chi_\text{A}(z) \rd z,
\end{equation} 
where $H$ is the pore width.
Moreover, we explore  the brush height, $\langle h_\text{b}^*\rangle$, defined as in~\cite{h1,dimitrov2}, 
\begin{equation} \label{eq:11} 
\langle h_\text{b}^*\rangle =2\frac{\int \rd z\, z\rho_\text{s} (z)}{\int \rd z \,\rho_\text{s} (z)}.
\end{equation}
\section{Results and discussion}\label{results}

\subsection{Parameters of the model}

In this work, without loss of generality, we employ solely one water-like model 
designated as W1 in~\cite{clark}. 
The model is characterized by the diameter of a particle, $\sigma=3.0342$~\AA,
the depth and range of the attractive potential, $\varepsilon/k=250$~K and $\lambda_\text{ff}$,
the association energy, $\varepsilon_\text{as}/k=1400$~K, and 
the cut-off distance of the attractive site-site potential, $r_\text{c}=2.10822$~\AA.

In order to avoid cumbersome notations, let us introduce the species subscripts $i,j$, 
referring to the brushes grafted at two walls, $i,j=1,2$, and to a fluid, if $i,j=3$.
Thus, according to the notation introduced in the description of the model, 
we deal with: $\sigma_{33}=\sigma$, $\varepsilon_{33}=\varepsilon$, 
$\lambda_{33}=\lambda_\text{ff}=1.7889$ (for W1 model).
As concerns the brush and cross brush-fluid interactions, we choose
$\sigma_\text{cc}=\sigma$, $\lambda_{ij}=\lambda_\text{bb}=1.4$, 
$\varepsilon_{ij}=\varepsilon_\text{bb}=\varepsilon$ for $i,j=11$, 12, 21 and 22.
Next, $\sigma_{i3}=\sigma_{3i}=\sigma$, 
$\varepsilon_{i3}=\varepsilon_{3i}=\varepsilon_\text{bf}=\varepsilon$ 
and $\lambda_{i3} = \lambda_{3i}=\lambda_\text{bf}=1.4$ for $i = 1$ or $2$.

The fluid parameters, $\sigma$ and $\varepsilon$ are chosen as the length and energy units,
respectively. The dimensionless quantities, such as diameters, pore width
and energy of interactions, are marked by an asterisk.
The dimensionless temperature, $T^*$, is defined as common,  $T^*=kT/\varepsilon$.

In all figures, we used the reduced temperature, $T^*_\text{r}$,  $T^*_\text{r}=T^*/T^*_\text{cb}$, where 
$T^*_\text{cb}$ is the critical temperature of the bulk water-like W1 fluid model ($T^*_\text{cb}=2.72$).
The energy of interaction of water-like species
with the pore walls is denoted as $\varepsilon^*_\text{fw}$. 
As it is  mentioned above, the segments of chains are just confined in the pore, but 
do not interact with the walls, i.e. $\varepsilon^*_\text{bw}=0$, throughout this study.

\subsection{Results}

\begin{figure}[!t]
\centering
\includegraphics[width=0.49\textwidth,clip]{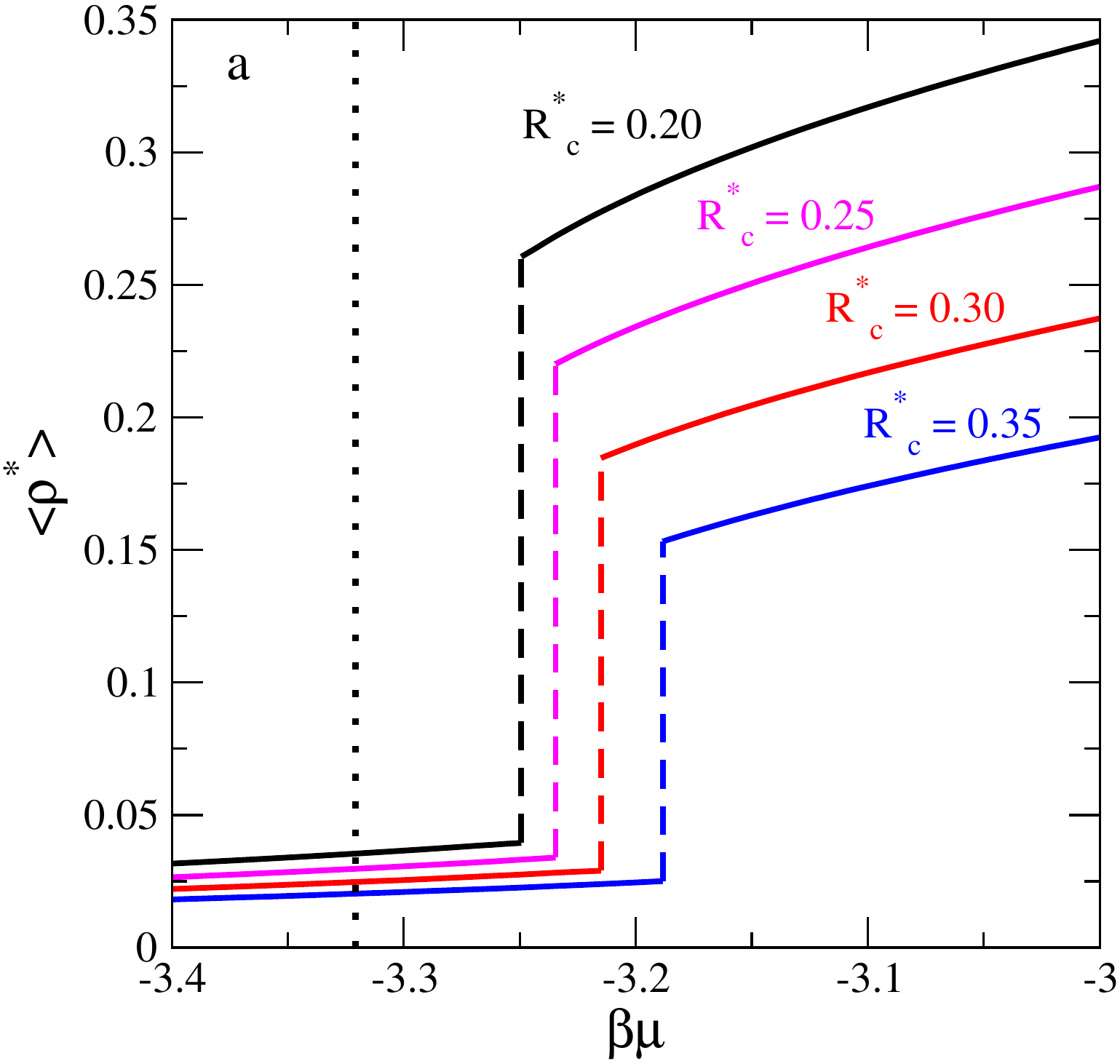}
\includegraphics[width=0.49\textwidth,clip]{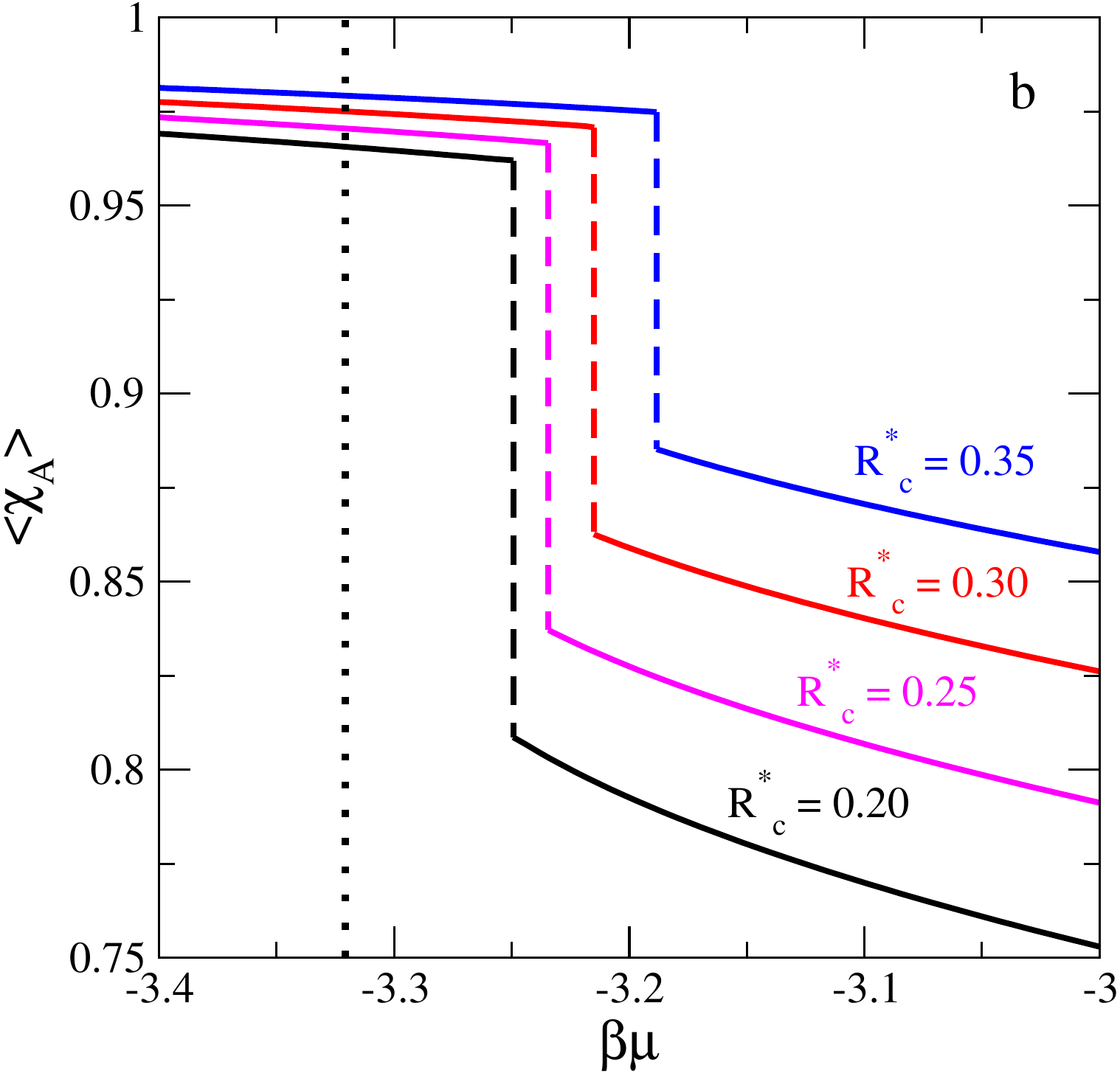}
\includegraphics[width=0.49\textwidth,clip]{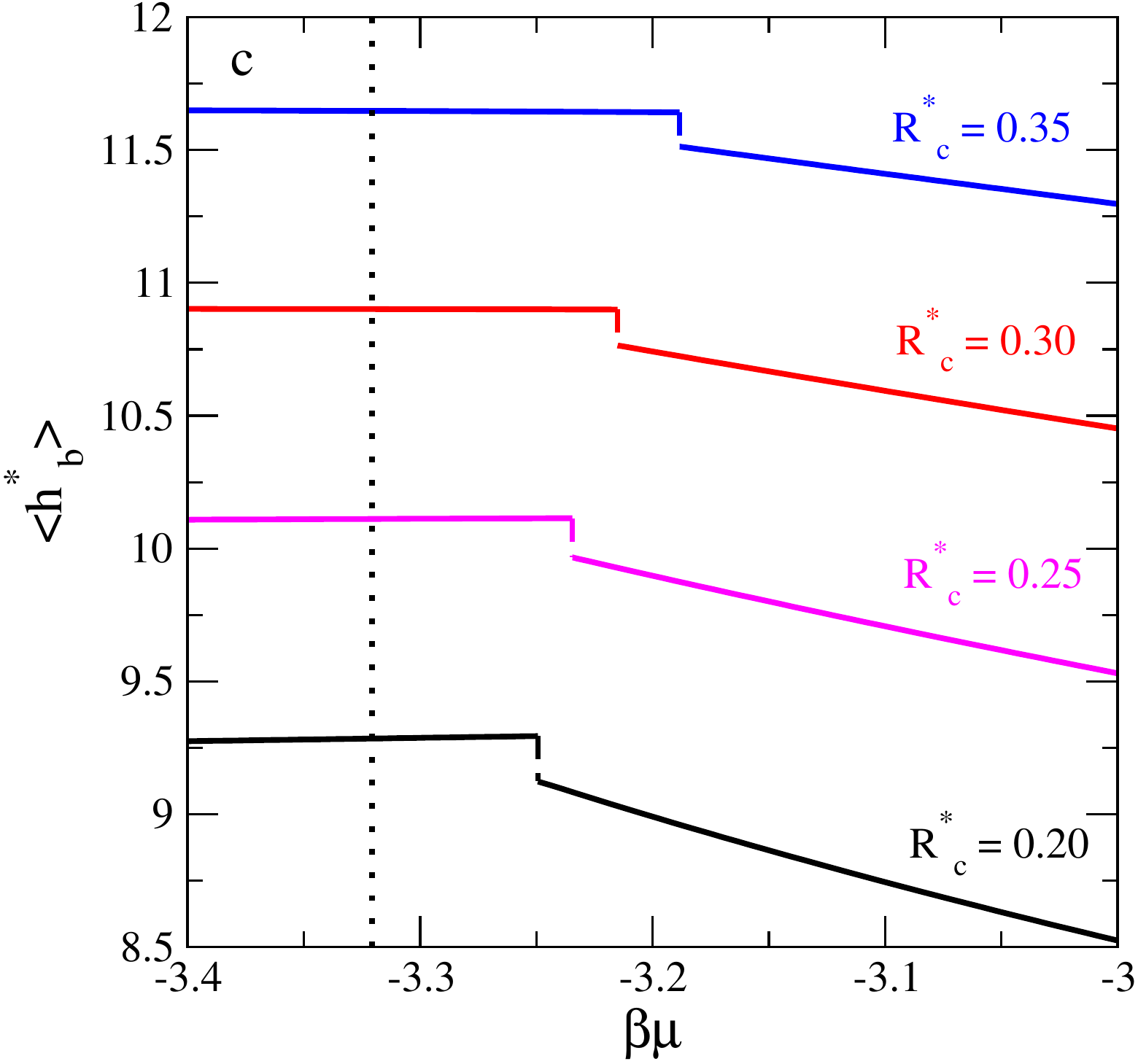}
\includegraphics[width=0.49\textwidth,clip]{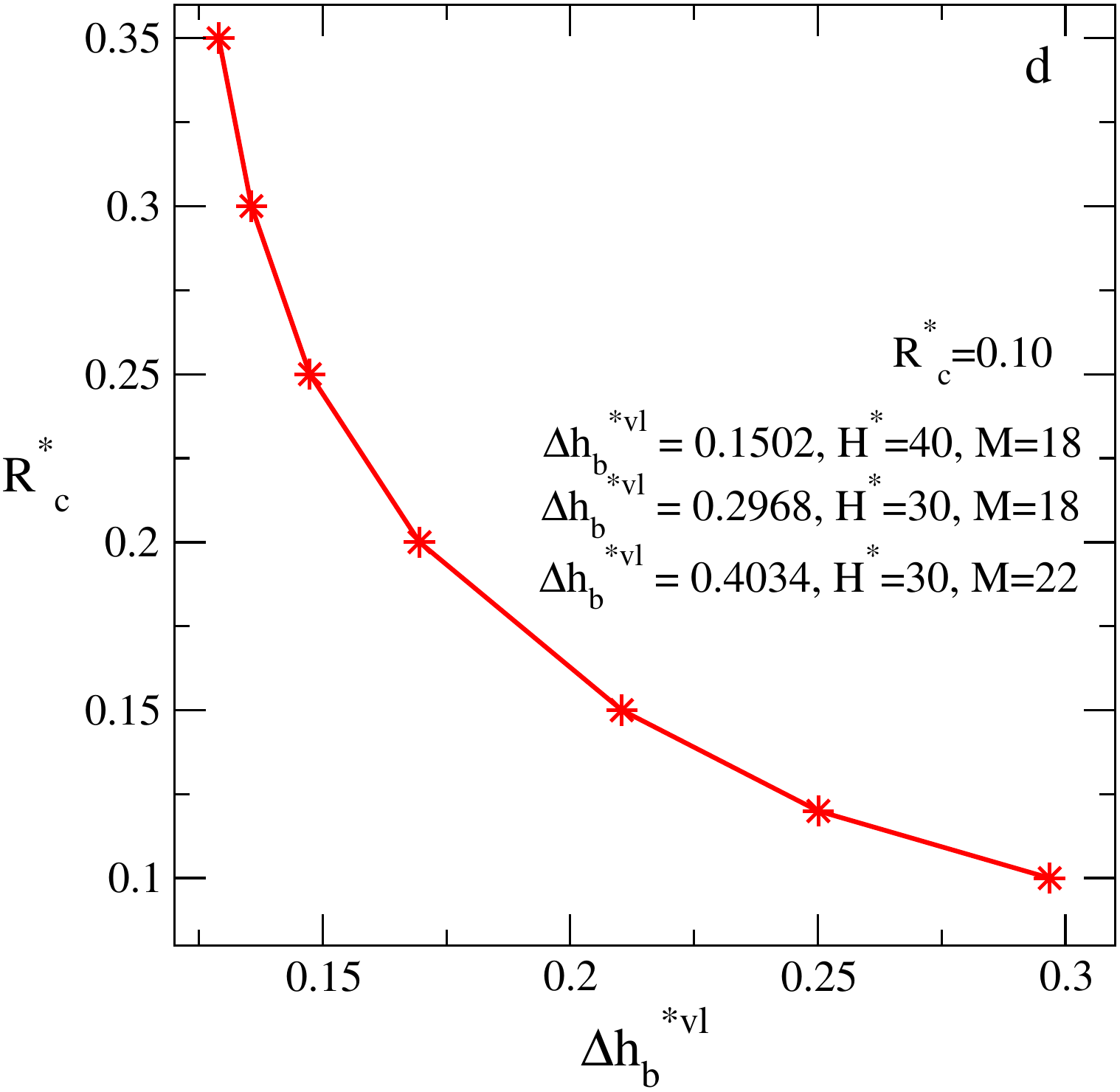}
\caption{ (Colour online)
Adsorption isotherms of the W1 water model in the slit-like pore of the width, $H^*=30$,
at temperature, $T_\text{r}^*= 0.85$, 
upon changing the grafted polymer layer density, $R^*_\text{c}$.
Panels (a), (b) and (c) refer to the projections, $\beta\mu$--$\langle \rho^* \rangle$,
$\beta \mu$--$\langle \chi_\text{A} \rangle$ and $\beta \mu$--$\langle h^*_\text{b} \rangle $, respectively ($M=18$).
Dotted line in panels (a), (b) and (c) corresponds to the chemical
potential at VL coexistence for the bulk fluid ($\beta \mu_0 =-3.3208$ at $T_\text{r}^*=0.85$).
Panel (d): Change of the brush height upon VL transition at different $R^*_\text{c}$.
In all panels: $\varepsilon^*_\text{fw} = 8.311$, $\varepsilon^*_\text{bb} =\varepsilon^*_\text{bf}=1$,
$\lambda_\text{bb}$=$\lambda_\text{fb}=1.4$. 
}
\label{FIGURE:1}
\end{figure}

The solution of  equations for the density profiles permits to construct the adsorption
isotherms and further discuss the phase diagrams.
A set of results reported in figure~\ref{FIGURE:1} was obtained at different 
brush density, $R^*_\text{c}$, for chains with $M =18$ in a quite wide pore $H^*=30$. 
The attraction  strength between fluid species and pore walls is at $\varepsilon^*_\text{fw} = 8.311$,
similar to the previous studies~\cite{trejos-x5}. The pore walls are rather strongly attractive
with this $\varepsilon^*_\text{fw}$.
The adsorption isotherms of the W1 water model in terms of average fluid density in the pore on 
chemical potential at a fixed $T_\text{r}^*=0.85$ are shown in panel (a) of figure~\ref{FIGURE:1}. 

The isotherms describe the vapour-liquid phase transition of a fluid in the pore and all of them
exhibit hysteresis. However, only  stable branches are plotted in the figure.
The magnitude of the average density jump decreases as the brush density, $R^*_\text{c}$,
increases. As was discussed in figure~1~(a) of  \cite{trejos-x5}, 
capillary condensation, $\mu < \mu_0$, in the pore
is observed in  the absence of  grafted chains on the walls whereas the capillary 
evaporation, $\mu > \mu_0$, is induced by the presence of grafted chains
($\mu_0$ refers to the chemical potential value at bulk coexistence).
The vapour density remains almost constant with the augmenting chemical potential up to the 
equilibrium phase transition
point in the pore. On the other hand, trends of the behaviour of liquid density after transition 
are determined by the brush density.
If the brush density is low, i.e., there is enough space in the pore for fluid adsorption, 
then the liquid density visibly grows upon increasing the chemical potential.
However, at a high brush density, e.g., $R_\text{c}^*=0.35$, the liquid density branch is less affected 
by the changes of chemical potential. In other words, the inclination of the liquid branch 
is determined by the $R_\text{c}^*$ value.

A set of projections of adsorption isotherms in terms of $\langle\chi_\text{A}\rangle$--$\beta \mu$ 
is shown in figure~\ref{FIGURE:1}~(b).
Here, the $\langle\chi_\text{A}\rangle$ values corresponding to vapour remain almost constant and close to unity 
upon increasing the chemical potential up to the phase transition point. 
Thus, the vapour phase at each $R_\text{c}^*$ is predominantly
composed of non-bonded molecules. The fraction of non-bonded particles in liquid phases is much lower.
The magnitude of the jump of $\langle\chi_\text{A}\rangle$ at the transition  decreases upon increasing $R^*_\text{c}$,
actually determined by the jump of the adsorbed fluid density.

Changes of the  height of the layer of tethered molecules upon the phase transition of the fluid 
in the pore are shown in terms of the average brush height, $\langle h^*_\text{b} \rangle$ on $\beta \mu$ in figure~\ref{FIGURE:1}~(c). 
The brush height jump is small upon the fluid phase transition.
The $\langle h_\text{b}^*\rangle$ is smaller  in the liquid phase compared to the vapour phase at each $R_\text{c}^*$.
In the liquid phase, the average brush height decreases upon increasing the adsorbed fluid density.
However, the inclination of the lines depends on $R_\text{c}^*$. 
The polymer chains comprising the brush are much more extended at high values of $R_\text{c}^*$, in comparison 
with low  $R_\text{c}^*$. Thus, the conditions of adsorption gedankenexperiment affect the brush stiffness.
In order to appreciate the magnitude of the jump of $\langle h_\text{b}^*\rangle$ 
upon changing $R_\text{c}^*$, we have constructed a plot shown in panel (d) of figure~\ref{FIGURE:1}.

We observe that the change of the brush height upon VL transition 
($\Delta h_\text{b}^{*\text{vl}}= \langle h_\text{b}^{*\text{v}}\rangle - \langle h_\text{b}^{*\text{l}}\rangle $) decreases as the grafting density, $R^*_\text{c}$, increases.
A smaller jump of $\Delta h_\text{b}^{*\text{vl}}$ at a high $R^*_\text{c}$ is due to a smaller available volume 
for adsorption of fluid molecules and, consequently, to a weaker influence of fluid species on 
the elongation of the chains in a dense brush.
Changes of the brush height upon VLE transition in the adsorbed fluid are most pronounced
at  quite low values of $R_\text{c}^*$. One example concerns $R_\text{c}^*=0.1$.
We performed additional calculations for the same conditions ($R_\text{c}^*=0.1$, $M=18$), 
but in a wider pore $H^*=40$. Then, $\Delta h_\text{b}^{*\text{vl}}$ changes from 0.2968 for $H^*=30$ to 0.1502 
for $H^*=40$.
On the other hand, if we change the nominal change length from $M=18$ to $M=22$ at a fixed pore width,
$H^*=30$ and at $R_\text{c}^*=0.1$, then $\Delta h_\text{b}^{*\text{vl}}$ changes from 0.2968 to 0.4034 (see the inset in panel~(d) 
of figure~\ref{FIGURE:1}). Consequently, changes of brush elongation in terms of $\langle h_\text{b}^{*}\rangle$ are
sensitive to the pore width and to the length of the chains at a fixed temperature. This observation can have implications
for the setup of systems with controllable thermodynamics of adsorption and fluid transport through
nanochannels. 

\begin{figure}[!t]
\centering
\includegraphics[width=6.9cm,clip]{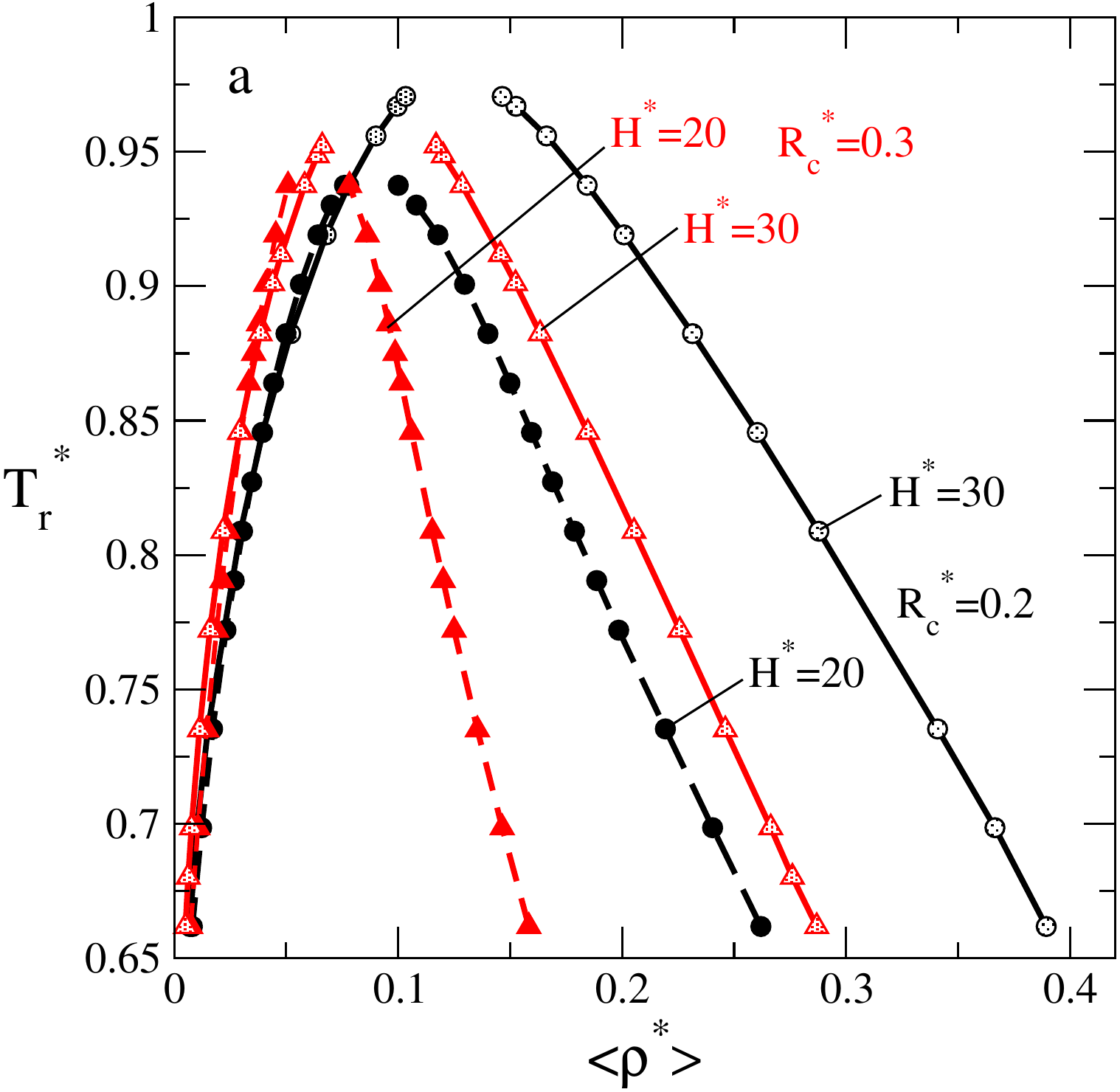}
\includegraphics[width=6.9cm,clip]{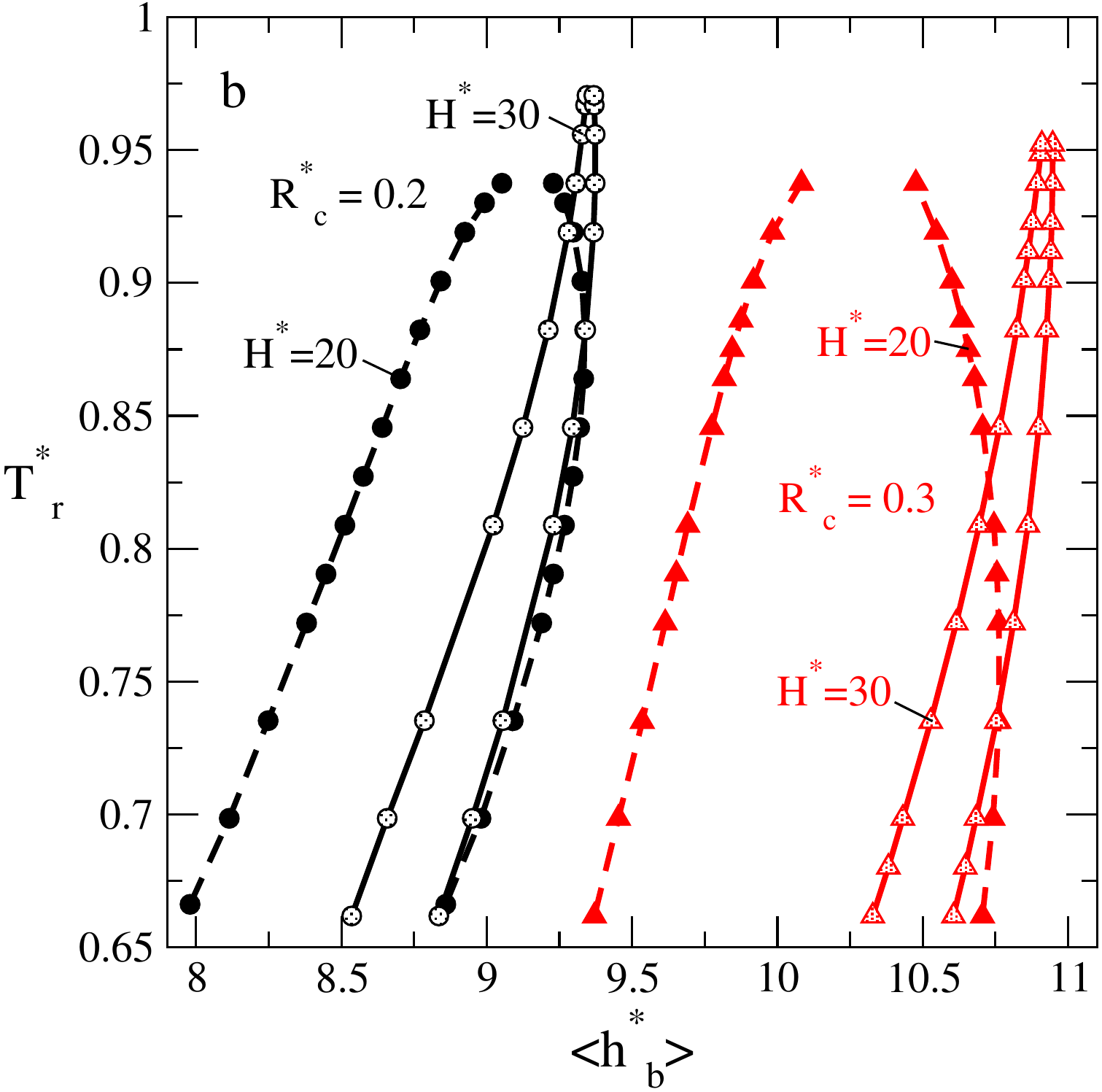}
\includegraphics[width=6.9cm,clip]{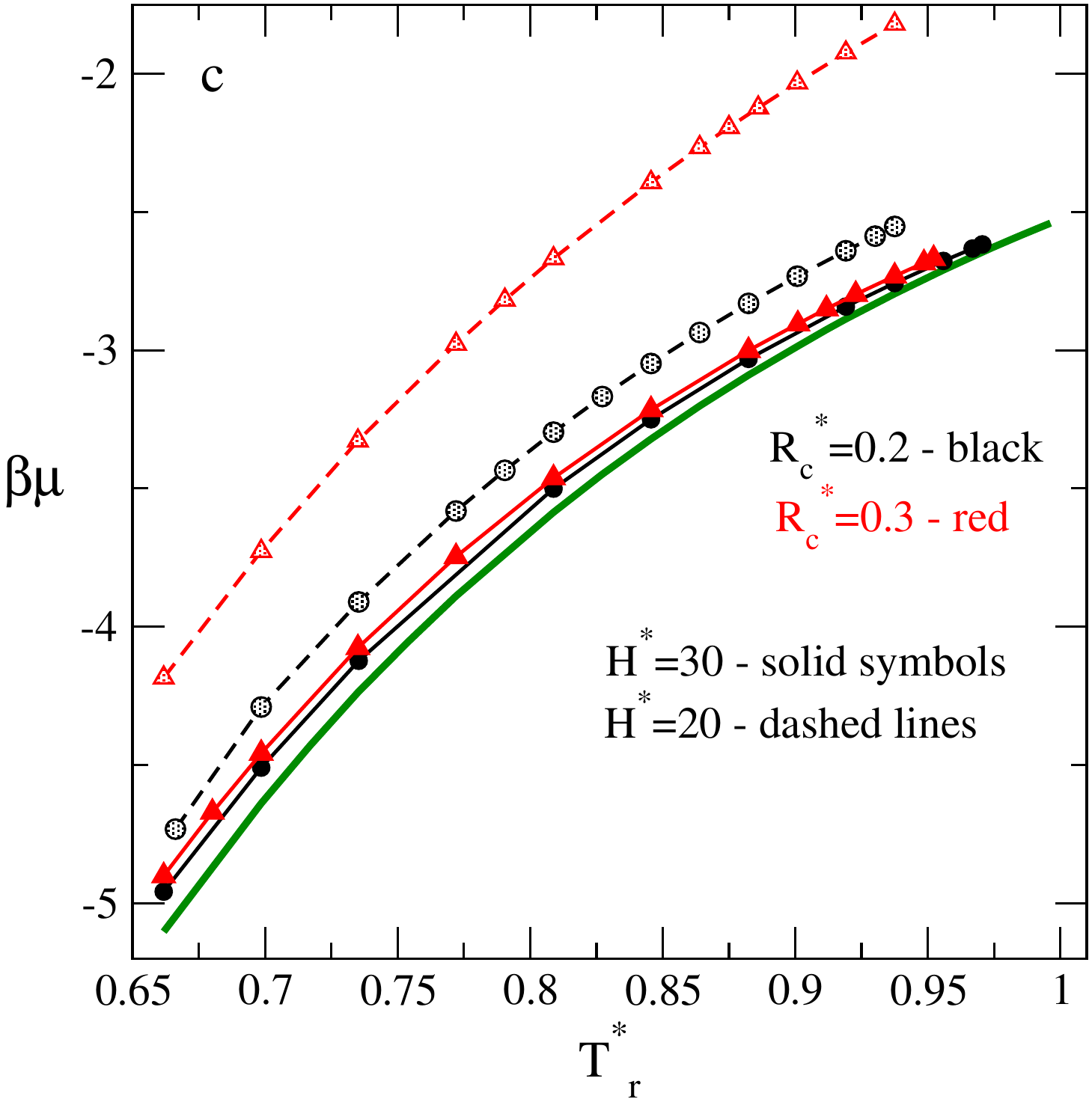}
\caption{ (Colour online) 
Projections of the VL coexistence envelope 
for W1 water model upon changing the tethered polymer layer density, $R^*_\text{c}$ and pore width, $H^*$. 
Panels (a), (b) and (c) refer to  $\langle \rho^* \rangle$--$T^*_\text{r}$,
 $\langle h^*_\text{b} \rangle$--$T^*_\text{r}$  and   $T^*_\text{r}$--$\beta \mu$ projections, respectively.
The nomenclature of lines and symbols is given in the figures.
The bulk fluid coexistence is plotted as a thick green line in panel (c). 
All the parameters are as in figure~\ref{FIGURE:1} [panels (a), (b) and (c)].
}
\label{FIGURE:2}
\end{figure}

After evaluating various adsorption isotherms, we can construct the VL coexistence envelopes. 
In figure~\ref{FIGURE:2}~(a), the effect of $R_\text{c}^*$ on the  $\langle \rho^* \rangle {-} T^*_\text{r}$ projection 
of the coexistence at two values of the pore widths $H=20$, and $H=30$, is displayed.
As concerns the effect of $R_\text{c}^*$ at constant $H^*$, we note that the
coexistence   envelope shrinks with increasing $R_\text{c}^*$.
The liquid branch is much more affected by the value of $R_\text{c}^*$ compared to the vapour branch density.
The effect of $H^*$ at constant $R^*_\text{c}$ can be summarized as follows.
Upon increasing $H^*$, the phase diagram in the average density-temperature plane widens, 
principally due  to the augmenting liquid density at coexistence.
In general, shrinking of the coexistence envelope is observed upon the augmenting confinement
due to a decreasing $H^*$ or increasing $R_\text{c}^*$. In both cases, the critical temperature decreases
very slightly. Still, it is close to the bulk critical temperature.
The critical density decreases as well. Apparent difference of the
critical density in the pores and in the bulk, as well as of the
densities along coexistence, can be attributed in part to 
the normalization of $\langle\rho^*\rangle$ using $H^*$, in contrast to a common normalization of the
adsorbed density through the available volume for adsorption in the theory
of fluids in disordered porous media, see e.g.,~\cite{orest,jps}.


\enlargethispage{2pt}

In order to explore the effect of $R_\text{c}^*$ and $H^*$ on the brush height, 
the  $\langle h_\text{b}^* \rangle {-} T^*_\text{r}$ projections of the coexistence envelope are shown in figure~\ref{FIGURE:2}~(b).
At a fixed $R^*_\text{c}$, either at 0.2 or 0.3, if the pore width, $H^*$, increases from $H^*=20$ to  $H^*=30$, 
the coexistence envelope changes from the thumb-like shape into ice-pick-like shape. 
A higher value of $\langle\rho^*\rangle$ along the liquid brunch of the coexistence in a pore $H^*=30$ 
compared to $H^*=20$ (cf. panel a of this figure),
means that the fluid particles are confined to a smaller volume and, as a consequence, the $\langle h_\text{b}^* \rangle$
value is higher in the pore $H^*=30$ compared to $H^*=20$. In other words, the brush, in contact with the 
liquid, is more ``compressed'' in a narrower pore compared to a wider pore. The effect of $R^*_\text{c}$ at
constant $H^*$ can be easily interpreted as well. A denser brush in terms of $R^*_\text{c}$ leads to a higher
$\langle h_\text{b}^* \rangle$ due to augmenting effects of excluded volume for 
the inner parts of chains corresponding to the brush body. 

Finally, we  plotted the $T^*_\text{r}{-} \beta \mu$ projection of the coexistence envelopes. 
The curves shown in figure~\ref{FIGURE:2}~(c) describe the capillary evaporation in the
entire temperature interval. The chemical potential at transition in 
the pore  becomes closer to the bulk coexistence if the pore width increases.
Moreover, in the wide pore, $H^*=30$, the transformation of vapour into liquid occurs at
a very similar  value of the chemical potential for two values of $R_\text{c}^*$, namely at 0.2 and 0.3.
In a narrower pore, $H^*=20$, though the difference of the values of chemical potential at transition is
substantial.

An overall thermodynamic picture described in the previous figures follows from the distribution
of species in the pore and in the resulting interface between brush and fluid. Therefore,
it is of importance to discuss the density profiles under different conditions.
The density profiles of grafted polymer layers and of a fluid, related to the present study, were
reported previously in several works. It is well known from the self-consistent field theory
that the system of grafted chains at moderate and high coverage is characterized by 
the density profile of parabolic shape~\cite{milner}, in agreement with findings by other methods
and experimental observations.  On the other hand, the density
profiles for grafted species were obtained using computer simulations (in the framework
of molecular dynamics and dissipative particle dynamics techniques)~\cite{malfreyt,dimitrov1,dimitrov2,dimitrov3}.
Specifically, the profiles reported from the laboratory of Binder refer to the model with
implicit solvent or to the model of a grafted layer in the presence of a hard sphere fluid only. Moreover,
the results concern either  a single flat substrate or  cylindrical pores of various diameters. 
The slit-like pore setup was considered in \cite{malfreyt}. We are not aware of simulations
data for systems that involve fluids with attractive interactions, either non-associative or
with chemical association. Therefore, the density profiles obtained
by using DF methodology are discussed with respect to computer simulations results at the level
of trends observed within both methods. 


\begin{figure}[!b]
\centering
\includegraphics[width=6.5cm,clip]{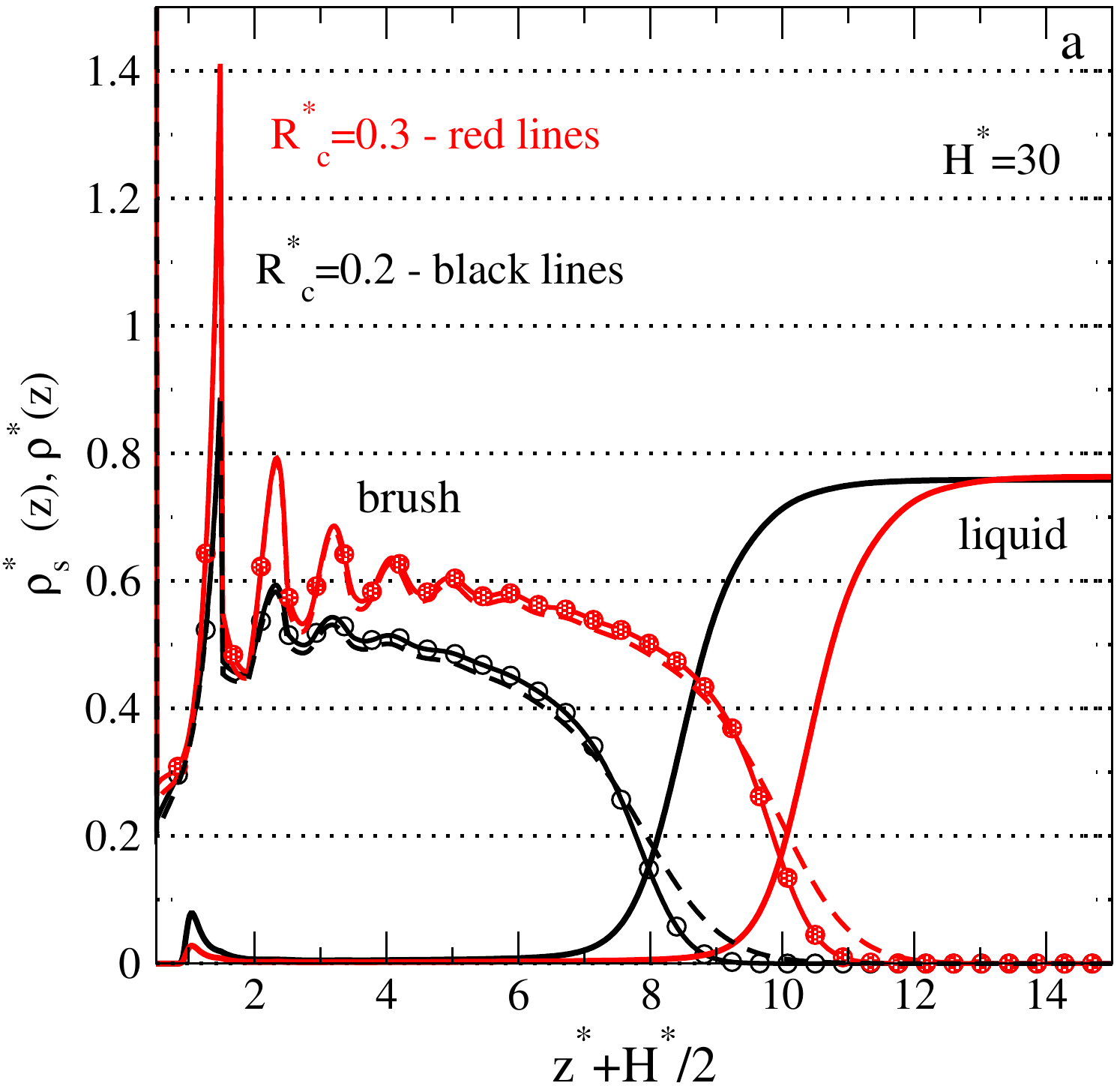}
\includegraphics[width=6.5cm,clip]{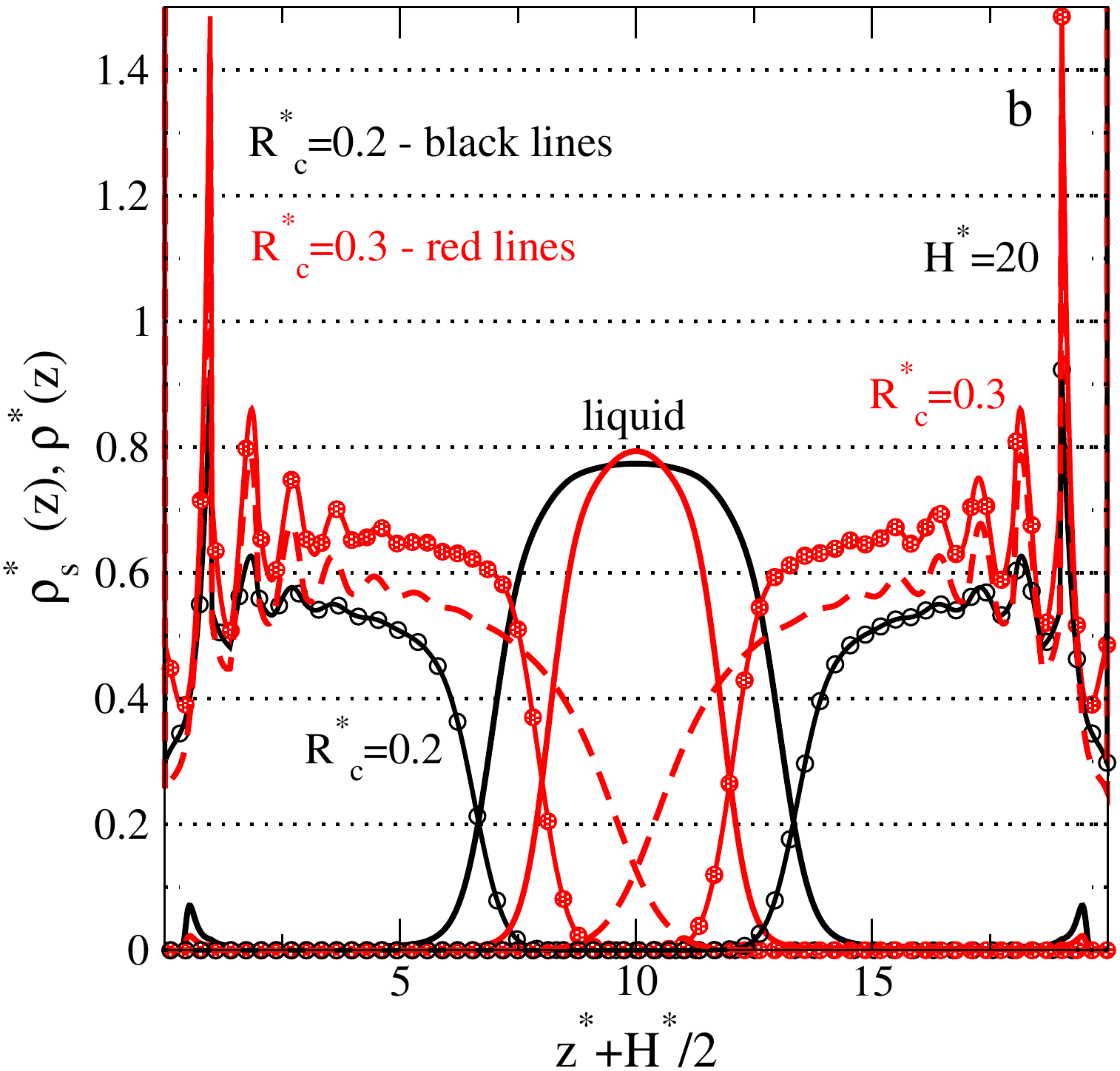}
\includegraphics[width=6.5cm,clip]{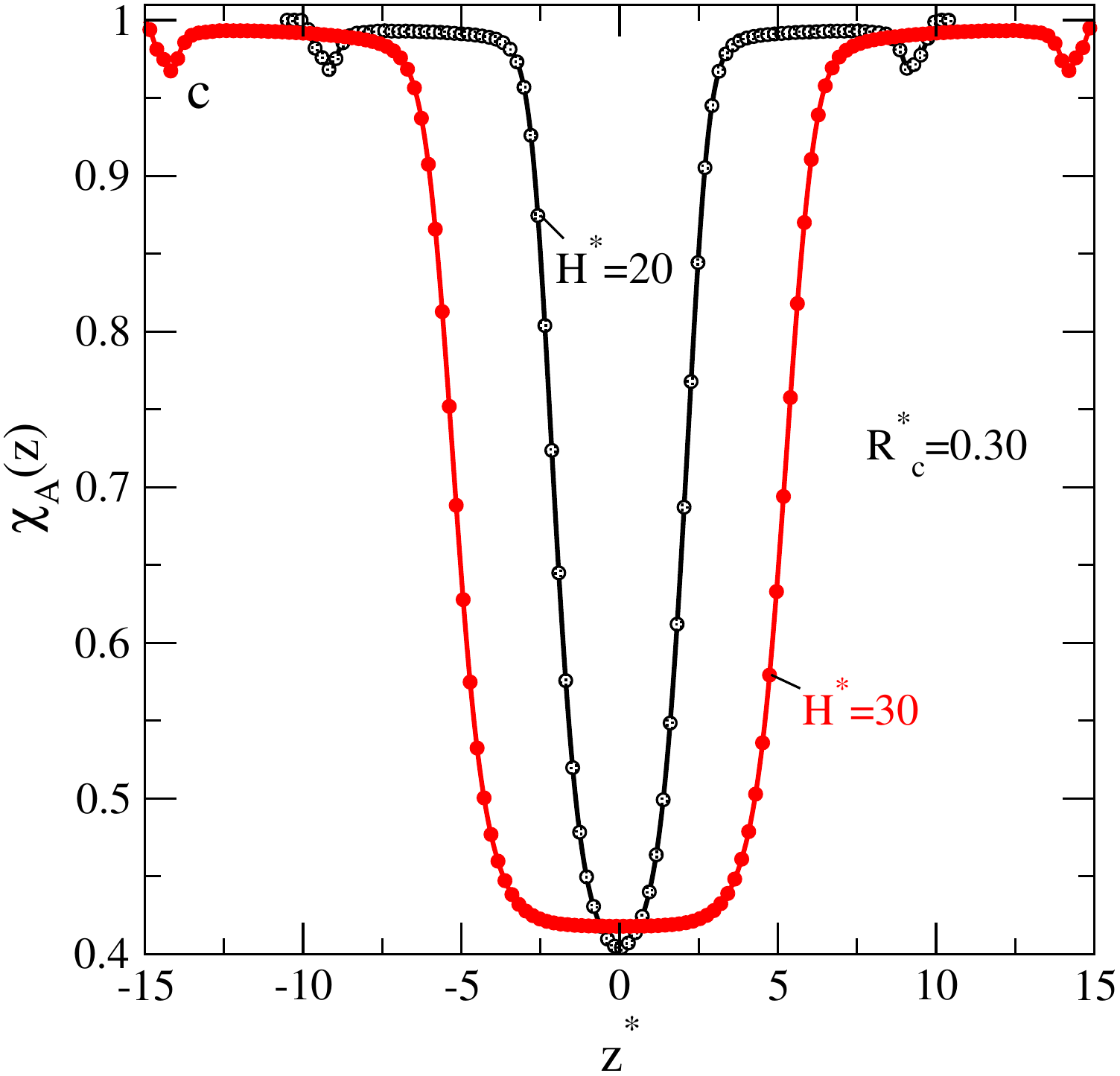}
\includegraphics[width=6.5cm,clip]{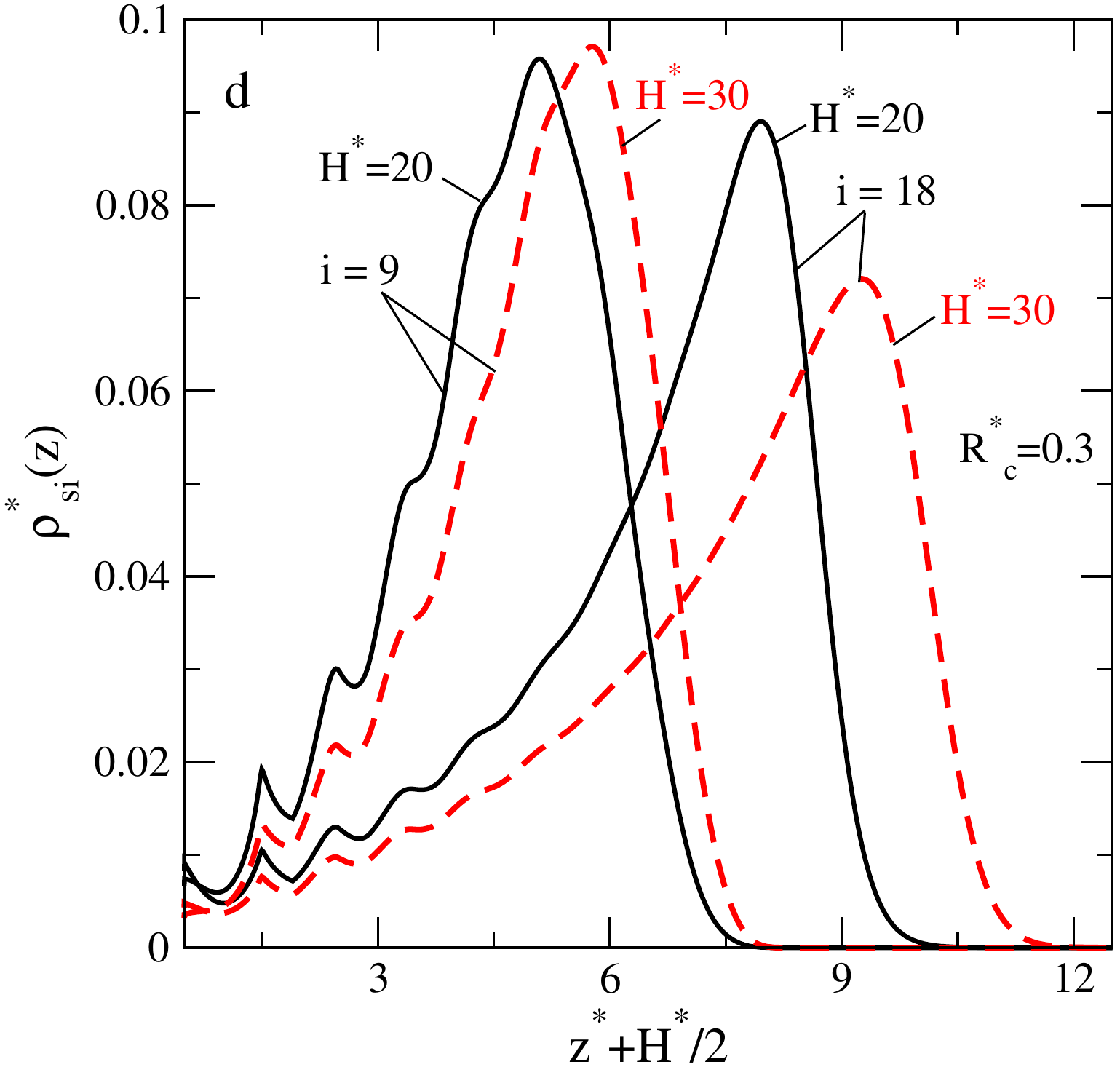}
\caption{(Colour online)
Examples of the fluid density profiles and brush density profiles [panels (a) and (b)],
and the density profiles for the non-bonded fluid species [panel (c)]
of thermal tethered polymer layer and of W1 fluid model
at VL coexistence for  $R^*_\text{c}= 0.2$ and 0.3.
Panel (d): The density profiles of selected segments $i$ (marked in the figure)
of grafted chains at fluid VL coexistence.
The dashed lines in panels (a) and (b) represent the brush profiles in
contact with fluid vapour at VL coexistence.
The pore width and the number of segments in grafted chain
molecules are $H^* = 20$ and 30, $M = 18$, $T^*_\text{r}=0.735$.
All other parameters are as in figure~\ref{FIGURE:1} [panels (a)--(c)].
}
\label{FIGURE:3}
\end{figure}

Examples of the density profiles for fluid species and for the brush at vapour-liquid
coexistence for fluid species are presented in figure~\ref{FIGURE:3}. 
The fluid density profile in the panels~(a) and (b) of this plot 
corresponds to a liquid state. Only one part of the
profiles is shown due to the symmetry of distributions w.r.t. pore center in panel~(a).
From  figure~\ref{FIGURE:3}~(a), we learn that the fluid density in the center 
of a wide pore ($H^*=30$) is practically constant and takes the value very close to the density 
of the bulk W1 water model at coexistence for the temperature in question ($\rho^*_\text{b}=0.7716$).
Upon increasing the surface coverage of grafted chains, $R_\text{c}^*$, from 0.2 to 0.3, the brush
layer becomes wider, indicating the stretching of chains. This behaviour is in accordance with what is observed
from the analysis of computer simulation results, see for example panels (c) and (e) of figure~5 from
\cite{dimitrov3}. A similar trend of behaviour with increasing the surface coverage is illustrated
in figure~4 of \cite{malfreyt}. Augmenting $R_\text{c}^*$ leads to a well packed structure of the inner part
of the grafted chains layer. A part of the density profile,  $\rho^*_\text{s}(z)$, describing the
external part of the polymer layer exposed to the adsorbed liquid decays similarly for two
values of the surface coverage considered in  figure~\ref{FIGURE:3}~(a).
Apparently, the stretching of chains is a quite strong effect,  
because the fluid occupies less space in the central part of
the pore at $R_\text{c}^*=0.3$ compared to $R_\text{c}^*=0.2$ [figure~\ref{FIGURE:3}~(a)]. 
Consequently, the average fluid density in the pore is lower in the former case compared 
to the latter, in accordance with the phase coexistence envelopes in figure~\ref{FIGURE:2}~(a).  
Moreover,  the profiles of brushes in figure~\ref{FIGURE:3}~(a) exhibit a collapse effect upon the fluid transition
from vapour to liquid, cf. dashed (brush in contact with vapour) and solid lines in figure~\ref{FIGURE:3}~(a),
in accordance with the corresponding phase diagrams. This effect is intrinsically out
of reach for the model considered in simulations in \cite{dimitrov1}.

In general terms, the shape of the interface between brush and liquid in a pore
with $R_\text{c}^*=0.2$ and $R_\text{c}^*=0.3$, as well as its width are similar. The fluid species in fact permeate 
only a small outer part of each brush. In contrast to this outer part,  
where the brush density profile decays, the inner part exhibits oscillations that develop
with increasing $R_\text{c}^*$. These oscillations witness a dense packing of brush segments.
A small probability to find fluid particles very close to
the pore wall is nevertheless observed as a result of fluid-wall attraction. 

Similar trends of behaviour of the profiles can be seen in figure~\ref{FIGURE:3}~(b) for the
case of a narrower pore, $H^*=20$. Apparently, the interface width is smaller. The inner 
parts of the brush profiles become flatter compared to panel (a), indicating a higher compression
of the brushes in this pore ($H^*=20$) in comparison with $H^*=30$.
This behaviour has its origin in the strong collapse of the brush layers 
at two walls upon the fluid transition from vapour to liquid. The brushes in contact with vapour
[dashed lines in figure~\ref{FIGURE:3}~(b)] exhibit a high degree of interdigitation, see \cite{trejos-x5} 
for a detailed description of this phenomenon within DF approach. By contrast, the liquid 
phase is dense and vigorously  separates brushes from two walls. The magnitude of the
observed changes depends on the pore width and grafted layer density. As concerns 
this kind of trends described by computer simulations, we refer to figure~4 of~\cite{malfreyt},
where the interdigitation of brushes is the result of augmenting density of grafted chains.

The effect of $H^*$ on the density profile of fluid species non-bonded at a site, $\chi_\text{A}(z)$, 
is shown in figure~\ref{FIGURE:3}~(c).
The fraction of non-bonded particles, $\chi_\text{A}(z)$,
is low in the central part of the pore due to a high fluid density there.
The density profile, $\chi_\text{A}(z)$,  shrinks as the pore becomes narrower. 
The width  of interface in terms of $\chi_\text{A}(z)$ follows the trends of the fluid density in this part of the 
pore.


An additional insight into the mechanism of ``compression'' of the brush upon changing the pore width
is given in figure~\ref{FIGURE:3}~(d). 
We show the changes of the  density profiles of selected segments, $\rho^*_{\text{s}i}(z)$, 
of chain molecules comprising the brush. The inner segments, $i<9$, behave similarly
in both cases, $H^*=20$ and $H^*=30$. The density profile of a middle segment, $i=9$,
exhibits a shift to the wall for $H^*=20$ compared to $H^*=30$, still the height of
the maximum is almost equal in both cases. More pronounced changes occur for the
density profiles of segments exposed to the liquid phase. In particular, the density
profile of the terminating segment, $i=18$, is much closer to the pore wall in the
case $H^*=20$ in comparison with $H^*=30$. Moreover, the maximum of this profile is
much higher for $H^*=20$, if we compare with $H^*=30$.

\begin{figure}[!t]
\centering
\includegraphics[width=.49\textwidth,clip]{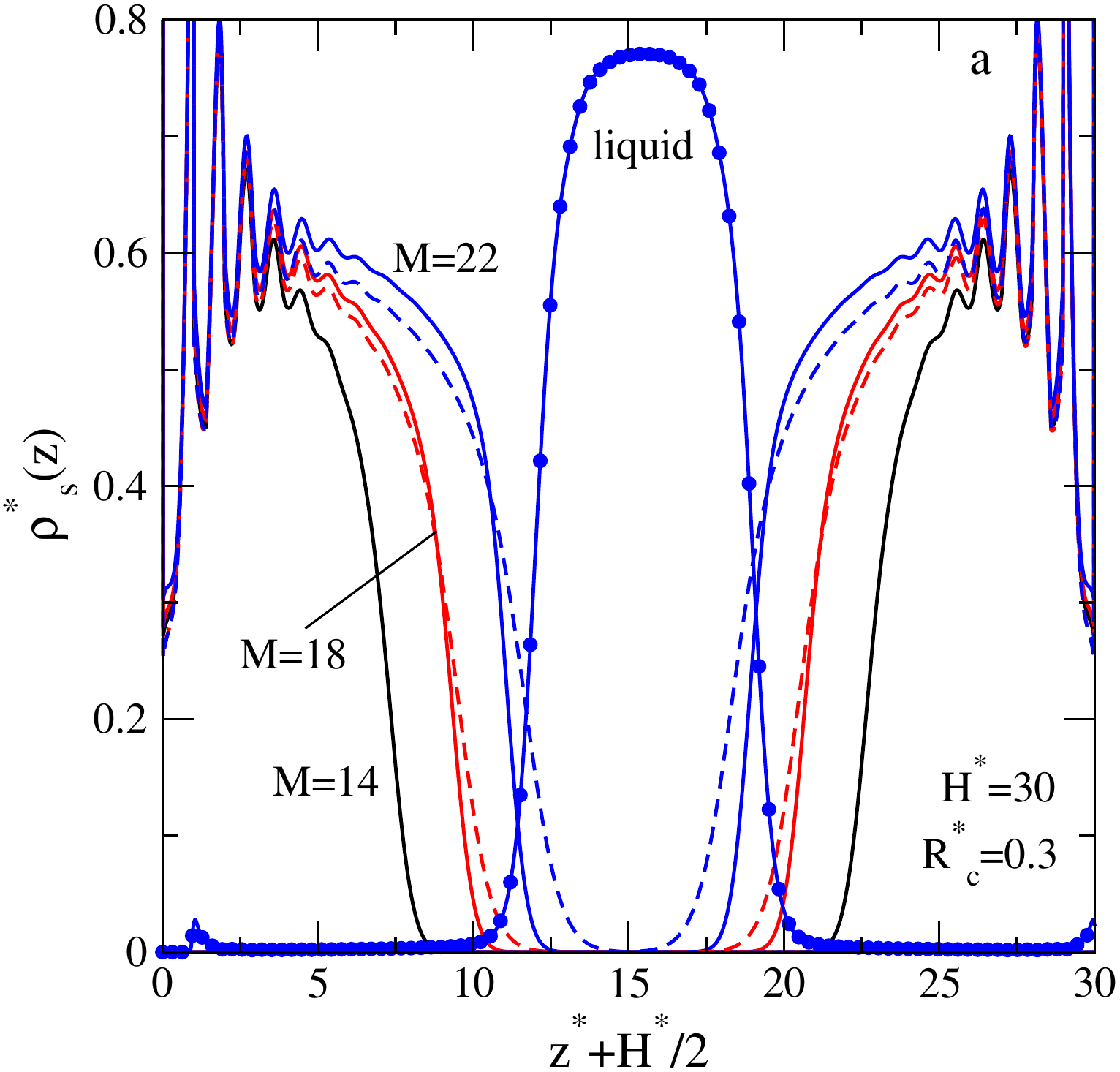}
\includegraphics[width=.49\textwidth,clip]{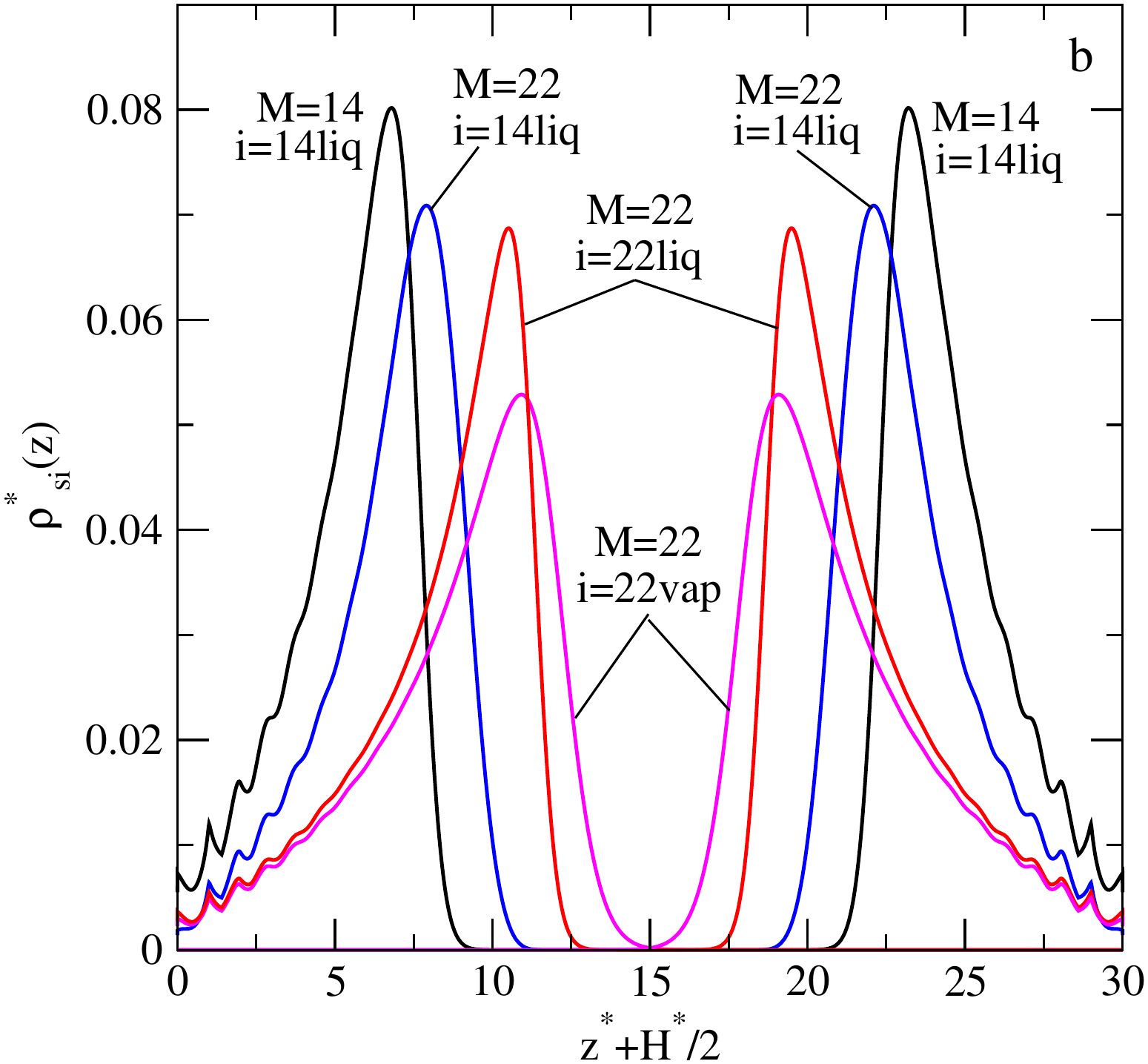}
\caption{(Colour online)
Effect of grafted chains length:
examples of the brush density profiles and of the fluid
distribution (for the model with $M=22$) in the pore [panel (a)],
and the density profiles of selected segments $i$ (marked in the figure)
of grafted chains at fluid VL coexistence.
The dashed lines in panel (a) represent the brush profiles in
contact with fluid vapour at VL coexistence.
The segments in contact with fluid vapour or liquid in panel (b) are marked as
ivap and iliq, respectively.
The pore width is $H^* = 30$  and the grafting density is at $R^*_\text{c}=0.3$,
$T^*_\text{r}=0.735$.
The number of segments in grafted chain is given in the figure.
All other parameters are as in figure~\ref{FIGURE:1} [panels (a)--(c)].
}
\label{FIGURE:3A}
\end{figure}

Our discussion of the density profiles of species with respect to the conclusions
reached by computer simulations of similar class of models would be incomplete
without describing the changes of the profiles on the number of segments of grafted 
chains. Some insights into this issue are provided in figure~\ref{FIGURE:3A}.  Namely, one can observe
that the height of the brush layer increases with increasing the chain length, provided 
the grafting density is kept constant, figure~\ref{FIGURE:3A}~(a). This behaviour from the DF theory 
is in agreement with the trends coming from computer simulations with
implicit solvent, cf. figure~5 of \cite{dimitrov3}. Again, shrinking of the
grafted polymer layers can be observed for models with different $M$ upon fluid phase
transition from vapour to liquid, figure~\ref{FIGURE:3A}~(a). Stretching of chains, at constant $R_\text{c}^*$, 
is more  pronounced with an increasing number of segments, cf. the density profiles
of the last segments for the models with $M=14$ and $M=22$, figure~\ref{FIGURE:3A}~(b). A similar
conclusion follows from simulations, see panels (b), (d) and (f) of figure~5 from~\cite{dimitrov3}. Moreover, essential changes of the profile describing the
last, most exposed to the fluid body, segments, occur upon condensation in the pore,
figure~\ref{FIGURE:3A}~(b). 

\begin{figure}[!b]
\centering
\includegraphics[width=0.49\textwidth,clip]{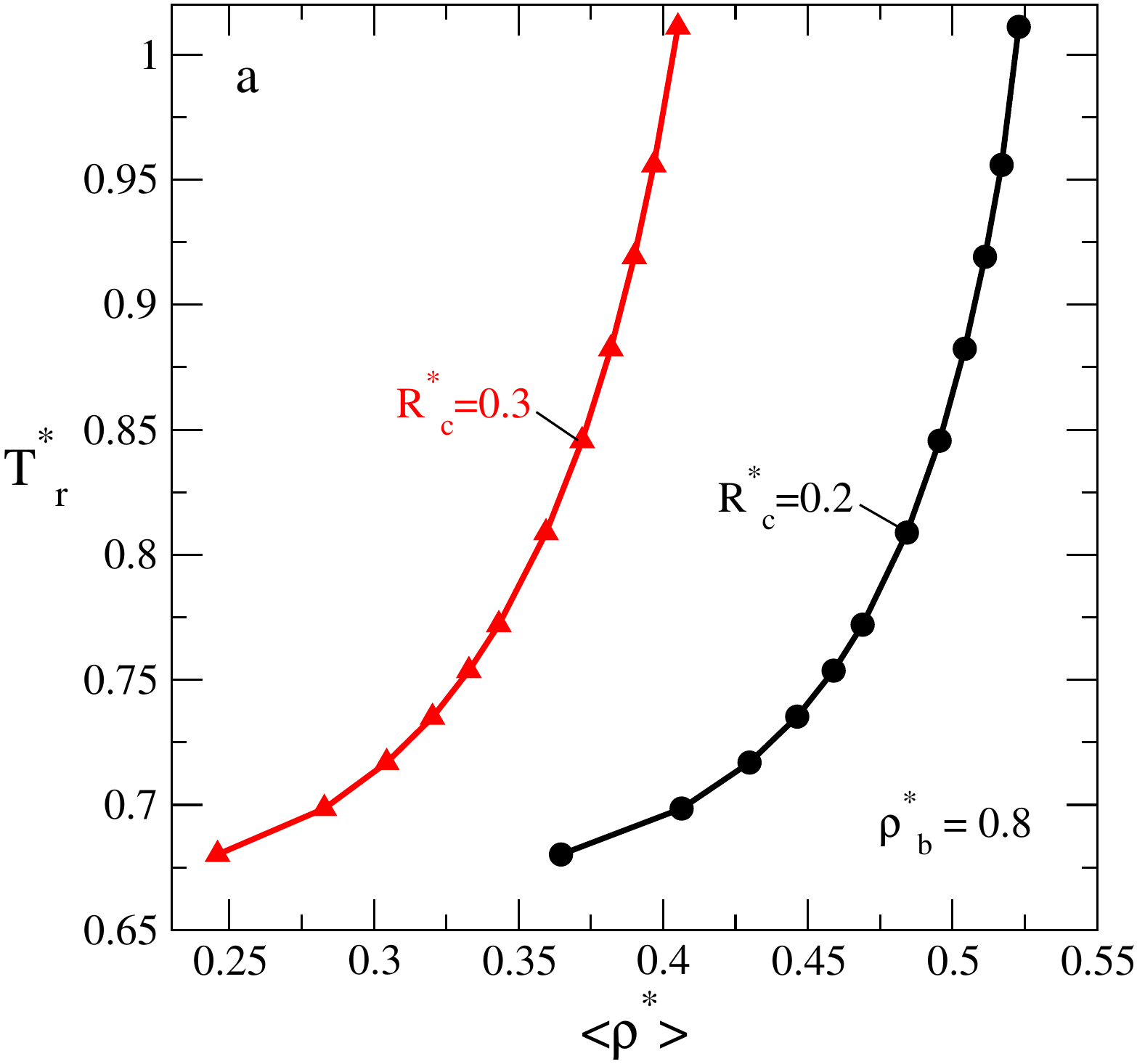}
\includegraphics[width=0.49\textwidth,clip]{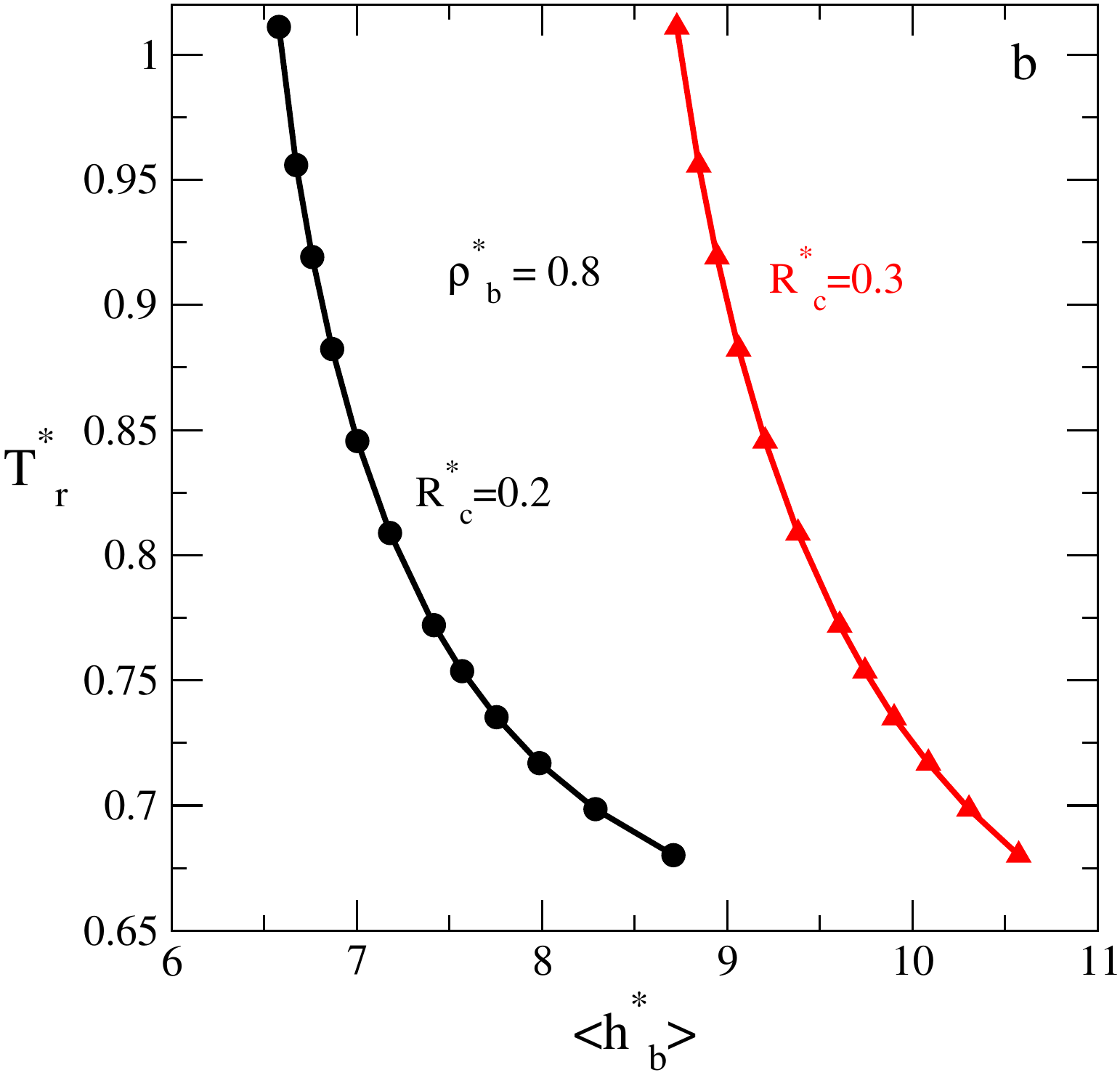}
\caption{(Colour online)
Density and brush height on temperature
at a fixed bulk liquid density, $\rho_\text{b}^*=0.8$, 
for W1 water model in a pore $H^* = 30$ ($M = 18$).
Panels (a) and (b) refer to $\langle \rho^* \rangle$--$T^*_\text{r}$
and  $\langle h^*_\text{b} \rangle$--$T^*_\text{r}$ projections, respectively.
All parameters are as in figure~\ref{FIGURE:1}.
}
\label{FIGURE:4}
\end{figure}

All previous observations concerning the thermal response of brushes are concerned
in one way or another with the vapour-liquid coexistence of the water-like fluid in the 
pores of different width and with different density of grafted polymer chains.
In the last part of this communication, we would like to use a slightly different setup.
Namely, we assume that the bulk fluid density is fixed at $\rho_\text{b}^*=0.8$. Then, one is
able to follow the temperature trends of behaviour of the systems in question in an
ample interval of temperatures starting for example from supercritical conditions 
down to $T^*_\text{r} \approx 0.65$ remaining in the bulk liquid phase. We restrict ourselves to 
the pore with the width $H^*=30$ only in this part of the work.

\begin{figure}[!b]
\centering
\includegraphics[width=7.2cm,clip]{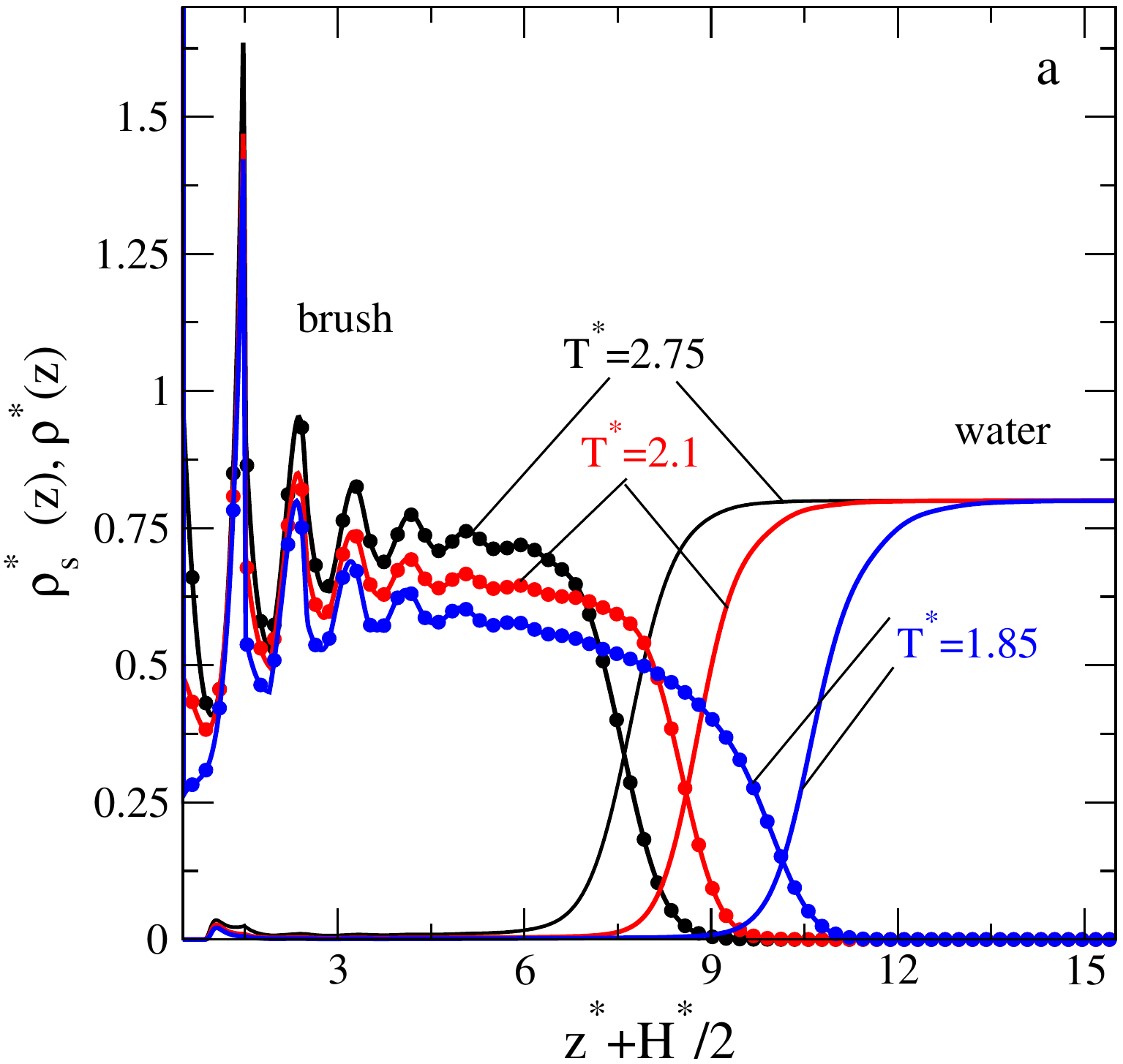}
\includegraphics[width=7cm,clip]{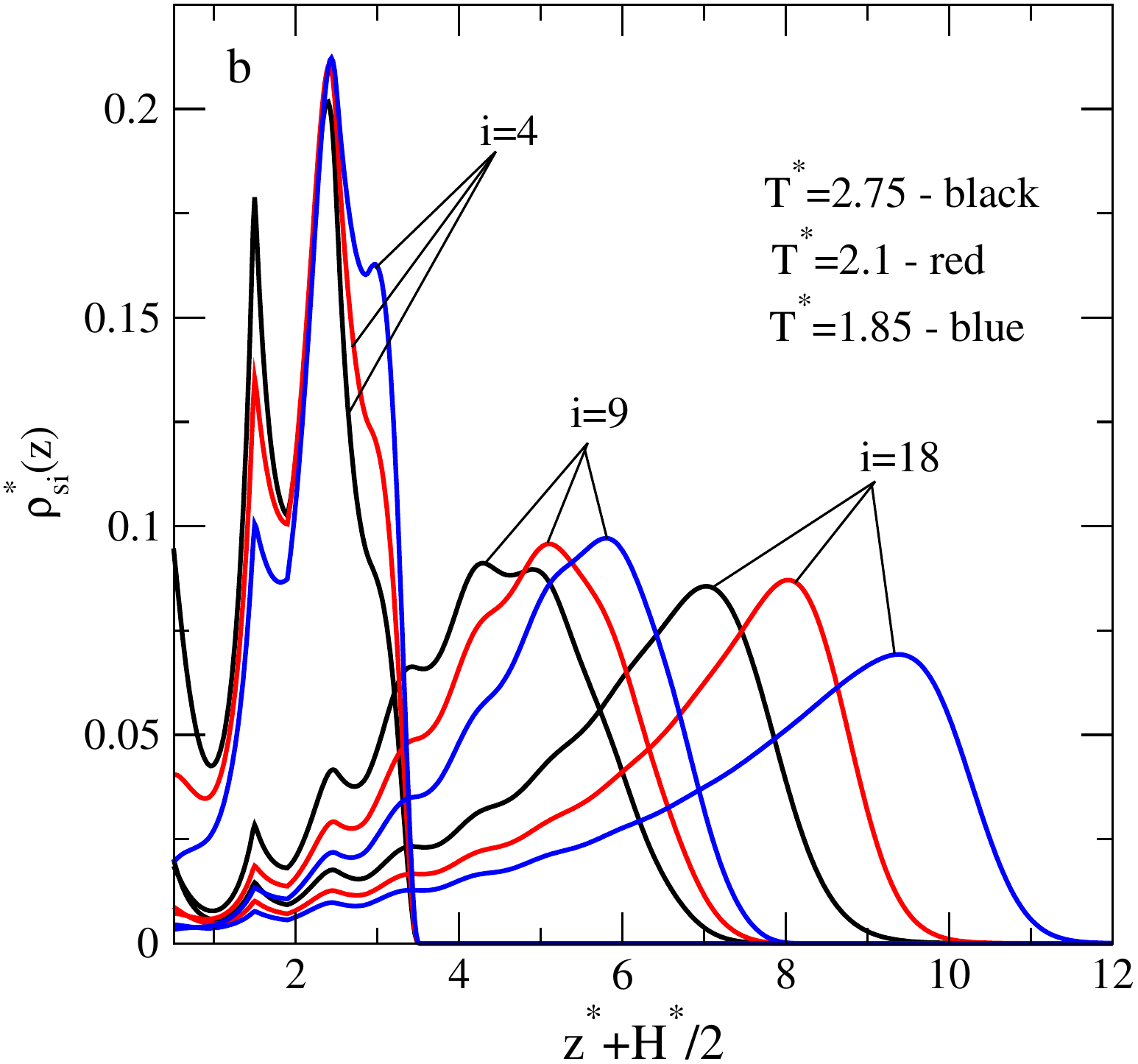}
\includegraphics[width=7cm,clip]{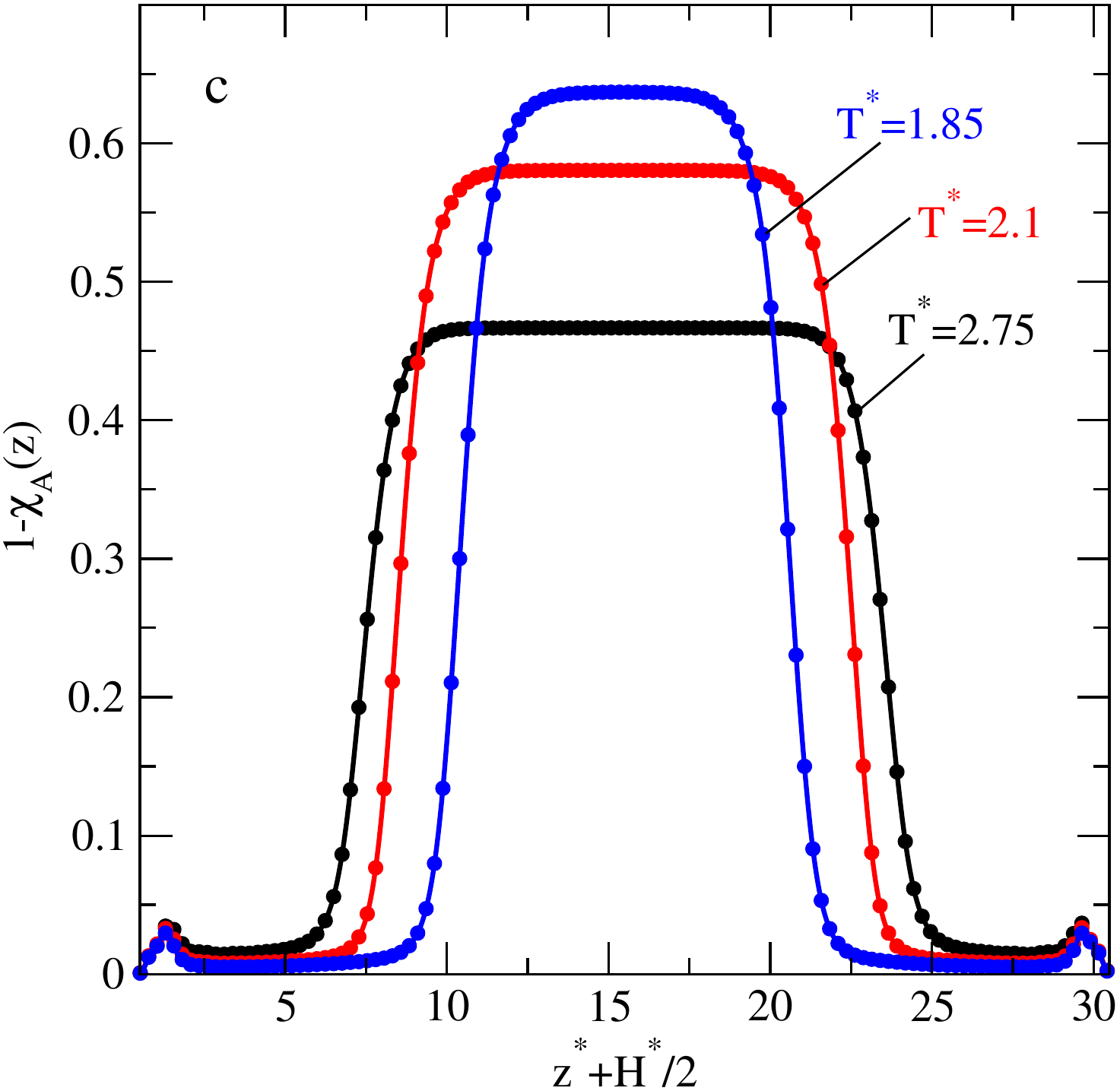}
\caption{(Colour online)
Examples of the liquid density profiles and brush density profiles [panel (a)],
density profiles of selected segments $i$ [panel (b)], and
density profiles of fluid species bonded at a site A [panel (c)] 
at different temperatures, $\rho_\text{b}^*=0.8$.
The pore width, the number of segments in grafted chain molecules  and 
the tethered polymer layer density are
$H^* = 30$,  $M = 18$ and $R^*_\text{c}= 0.3$, respectively. All other parameters are as in figure~\ref{FIGURE:1}.
}
\label{FIGURE:5}
\end{figure}

Evolution of the average fluid density in the pore with temperature, at a fixed bulk fluid density,
is shown in panel (a) of figure~\ref{FIGURE:4} for two systems with $R_\text{c}^*=0.2$ and with $R_\text{c}^*=0.3$.
In both cases, $\langle\rho^*\rangle$ monotonously decreases with a decreasing temperature. 
This behaviour is most pronounced  at low temperatures. 
Apparently, trends of behaviour of both curves in the figure are 
determined by the augmenting effects of attractive interactions while temperature
goes down. The effect of $R_\text{c}^*$ reduces only to the shift along the $\langle\rho^*\rangle$ axis.
The average brush height, $\langle h_\text{b}^* \rangle$, depends on temperature as well [figure~\ref{FIGURE:4}~(b)].
Namely, it increases with a decreasing temperature. This behaviour is straightforwardly 
correlated with the dependence of $\langle\rho^*\rangle(T_\text{r}^*)$. When the averaged adsorbed density
decreases, the average brush height increases. Two curves describing $\langle h_\text{b}^*\rangle(T_\text{r}^*)$ 
at different $R_\text{c}^*$ behave similarly. Just in the case $R_\text{c}^*=0.3$, a lower average density
of adsorbed fluid (in comparison with $R_\text{c}^*=0.2$) leads to higher values of the average brush height.       

It is of interest to find interpretation of this behaviour in terms of the microscopic structure
of the fluid and brush in the pore and the interface between them. One set of examples is 
presented in figure~\ref{FIGURE:5}.
From figure~\ref{FIGURE:5}~(a) we learn  that the brush density profile is essentially affected by temperature.
It is stiffer at high temperature and much more diffused at a low temperature. However, 
the fluid permeates the brush body better at a high temperature. Apparently, the liquid slab at
e.g., $T^*=2.75$  exerts a higher pressure over the brush due to a higher $\langle\rho^*\rangle$, cf. figure~\ref{FIGURE:4}~(a),
in comparison with e.g., $T^*=1.85$. 
Better insights into the changes of the brush height can be obtained from the analysis of 
the segment density profiles, figure~\ref{FIGURE:5}~(b), the inner part of the brush, i.e., for the segments
number $i<9$ does not change much upon a decreasing temperature. This is well seen for example
from the profile describing the segment $i=4$. The outer part of the brush described in terms
of the distribution of segments with $i>9$ is much more affected by the temperature changes.
The most pronounced effect is seen for the segment $i=18$ dangling quite freely in the fluid.

The structure of the liquid slab in the pore center can be analyzed in terms of the fluid density
profiles shown in panel (a) and in terms of the degree of bonding between water-like particles. 
The liquid situated in the pore center is predominantly composed of fluid particles participating in
bonds whereas water-like particles permeating the brush do not form bonds between them, figure~\ref{FIGURE:5}~(c).
How this distribution is manifested in the interface tension is an open question at the moment.
However, it is worth mentioning that the width of the interface in terms of $\chi_\text{A}(z)$
is sensitive to temperature as well.

\section{Summary and conclusions}\label{conclusions}

In this short report, we  presented theoretical results 
from a version of the DF theory concerning  adsorption of water into slit-like pores 
in  which the walls are chemically modified by grafted chain molecules that form molecular brushes.
The fluid model was taken from \cite{clark} and was incorporated into the DF approach.
The model for the entire system involves the non-electrostatic interactions solely.

In contrast to our previous research~\cite{trejos-x3,trejos-x4,trejos-x5}, we  included 
segment-segment attractive interaction in the form of  a square well. This modification
makes the brushes thermal on their own which leads to thermal responses due to fluid
species and to the grafted chains. Apparently, the theory formulated in the
spirit of semi-grand ensemble corresponds to the grafting from the procedure in the 
experiments because the bulk reservoir contains solely the fluid species.

The first setup employed here comprises the calculation of the thermodynamic potential
at a given external chemical potential and external field, and its minimization to obtain the density
profiles for adsorbed fluid and for grafted chains subsystem.  
We analysed and  discussed a few  examples of adsorption isotherms and of vapour-liquid coexistence envelopes 
for a confined fluid  using average density-temperature, chemical potential-temperature, 
chemical potential-brush height.  On the other hand,
we  explored the microscopic structure of the interface between the brush and adsorbed fluid.

Capillary evaporation of water-like fluid model is observed at a high density of tethered 
chains, $R^*_\text{c}$ in contrast to condensation in the absence of brushes in the pore.
Upon fluid vapour-liquid transition, the brush height decreases via the jump.
If the parameter $R^*_\text{c}$ that describes the brush density is high, the grafted polymer chains
attain an extended configuration leading to a high $\langle h^*_\text{b}\rangle$ for the fluid in the vapour phase
and a slightly smaller brush height. Under such a condition,  the fluid weakly permeates the brush.
A stronger brush collapse is expected at a lower tethered chains density.  

The brush density profiles at high values of $R_\text{c}^*$ have the shape that can be
interpreted as a well structured, dense inner part and a more flexible outer part
with dangling ends exposed to vapour or liquid. Upon vapour-liquid phase transition,
this outer part suffers most pronounced changes. In addition, it is important to note 
that the predictions concerned with the behaviour of the density profiles from the DF methodology 
are in agreement with the trends observed by using computer simulations 
as tools~\cite{malfreyt,dimitrov1,dimitrov2,dimitrov3}.
Complementary comparisons of the profiles from the
DF approach for fluid-grafted layer of short chains with Monte Carlo simulations 
were also presented (though for different models of interactions) in \cite{stah}.

On the other hand, we were interested in the exploration of temperature trends of behaviour 
of the system that are not related to the VL phase transition.
With this purpose, we  explored a setup at a constant bulk fluid density that permits to follow 
the thermal response of the system from the supercritical temperature down to  $T^*_\text{r} \approx 0.65$.

A decreasing temperature leads to a lower averaged density of adsorbed species. Consequently,
the average brush height monotonously increases. Swelling of the brush is due to the 
change of distribution of segments of grafted chains of the outer part of the brush. 

Finally, possible extensions or improvements of the model and method should be attempted.
As concerns the model, it would be of interest to investigate the effect of the block structure
of grafted polymers that would interact differently with  water-like particles.
Furthermore, the adsorption of mixtures is worth to explore, 
evidently the condensation and mixing properties would change due to the presence of brushes~\cite{kasia}.
In particular, if the demixing line is present in the phase diagram of the binary mixture, it
would shift and possibly change its inclination due to the presence of brushes on the walls.
In order to reach novel findings, however, one inevitably needs to employ  a more sophisticated modelling of the system
of grafted polymer molecules. Research along these lines is currently in progress in our laboratory 
and will be reported elsewhere.

\appendix

\section{Helmholtz free energy functional} \label{SuppMat}
The Helmholtz free energy is the sum of  ideal and excess terms,
$F=F_\text{id}+F_\text{ex}$, where the excess term is
expressed as the sum of the
contributions arising from different kinds of interactions in the system,
$F_\text{ex}=F_\text{hs}+F_\text{c}+F_\text{as}+F_\text{att}$, namely due to the volume exclusion, $F_\text{hs}$,
the connectivity of chains, $F_\text{c}$, due to association of fluid molecules,
$F_\text{as}$ and due to attractive van der Waals interactions, $F_\text{att}$. For the sake of brevity,
we  omitted the functional dependencies of all free energy terms.

The ideal part, $F_\text{id}$, is given by an exact expression~\cite{c40}
\[
\frac{F_\text{id}}{kT}= 
\int \!\!\rd{\bf R}\rho ^{(\text{c}1)}({\bf R})V_\text{b}^{(1)}({\bf R})/kT+\int \!\!\rd{\bf R}\rho ^{(\text{c}2)}({\bf R}
)V_\text{b}^{(2)}({\bf R})/kT
+\int \!\!\rd{\bf R}\rho ^{(\text{c}1)}({\bf R})\{\ln [\rho ^{(\text{c}1)}({\bf R%
})]-1\}
\]
\begin{equation}
 +\int \!\!\rd{\bf R}\rho ^{(\text{c}2)}({\bf R})\{\ln [\rho ^{(\text{c}2)}({\bf R%
})]-1\} + 
\int \!\!\rd\mathbf{r}\rho({\mathbf{r}})\{\ln [\rho (\mathbf{r})\Lambda^3] -1\}\;,\label{eq:id}
\end{equation}
where $\Lambda$ is thermal wavelength of fluid molecules.

The free energy due to hard-sphere interactions, $F_\text{hs}$, is
evaluated in the framework of  the White Bear version of the fundamental measure
theory \cite{ca2,c41}. This approach requires introduction of four scalar,
and two vector averaged densities. For the sake of brevity, the 
definitions of averaged densities are omitted, they are given by equations~(3)--(6) of \cite{c41}. 
Since the fundamental measures theory is well-known, we do not repeat it here. The definition for the
hard-sphere contribution to the free energy is given by equations~(1), (9), (10) and (11) of \cite{c41}.
Similarly, the connectivity contribution to the Helmholtz free energy, $F_\text{c}$,
was reported in several works,
see, e.g., equations~(13) and (14) of \cite{ca3}.

The term arising from associative interactions
between fluid molecules, $F_\text{as}$, results from the theory outlined 
in  \cite{trejos-x1,trejos-x2,YuWu2002}
\begin{equation}
F_\text{ass}/kT= 4\int \rd\mathbf{r}\, n_0(z) \zeta(z) \left\{\ln\chi_\text{A}(z)-\frac{1}{2}[\chi_\text{A}(z)-1]\right\},
\end{equation}
where $\chi_\text{A}(z)$ is the density profile of the fraction of molecules at 
position $z$ that are not bonded at the site~A,
\begin{equation}
\chi_\text{A}(z)=\left[1+n_0(z)\zeta(z)\sum_{\alpha \in \Gamma}\chi_{\alpha}(z)\Delta(z)\right]^{-1},
\end{equation}
and where $\Delta(z)$ describes the intermolecular site-site bonding,
$\Delta(z) = 4\piup K F g_\text{hs}(\sigma,n_i(z))$,
where $F = \exp(\varepsilon_\text{as}/kT)-1$.
The parameters of the model determine the expression for site-site bonding volume,~$K$~\cite{jackson1,jackson2},
\begin{align}
 K &=  \frac{\sigma^2}{72 d_\text{s}^{2}} \left\{
 \ln \left[ (r_\text{c} + 2d_\text{s})/\sigma \right]
 \left(6r_\text{c}^{3} + 18r_\text{c}^{2}  d_\text{s} - 24d_\text{s}^{3}\right)  
 \nonumber
 \right.
 \\ &
 + \left( r_\text{c} +2d_\text{s} - \sigma \right)
 \left. 
\left( 22d_\text{s}^{2} -5r_\text{c} d_\text{s} - 7d_\text{s} \sigma -8r_\text{c}^2 + r_\text{c} \sigma + \sigma^2 \right)
 \right\}.
 \end{align}
The contact value of the pair distribution function of hard spheres, $ g_\text{hs}(\sigma,n_i(\mathbf{r}))$, 
is calculated by using the equation given by Yu and Wu  \cite{YuWu2002}.   

The attractive interactions between all spherical species are described
in the framework of the mean field approximation. Assuming that interactions of 
fluid molecules with all the segments of chains 1 and 2 are identical, we have,
\begin{align}
 F_\text{att}&=\frac{1} {2} \int \rd {\bf r}_1 \rd {\bf r}_2 \rho(z_1) \rho(z_2) u_\text{att,ff}(r_{12})
 +\int \rd {\bf r}_1 \rd {\bf r}_2 \rho_\text{s}^{(1)}(z_1)\rho(z_2)u_\text{att,fc}(r_{12})\nonumber\\
 &+\int \rd {\bf r}_1 \rd {\bf r}_2 \rho_\text{s}^{(2)}(z_1)\rho(z_2)u_\text{att,fc}(r_{12}),
\end{align}
where $u_\text{att,ff}(r)$ and $u_\text{att,fc}(r)$ are the attractive parts of the
fluid-fluid and fluid-segment potentials, respectively.

\section{Density profile equations}

Density profile equations are obtained from equation~(\ref{15}). 
For fluid molecules, we obtain
\begin{equation}
 \rho(z)=\exp\left[\frac{\mu - \lambda(z)}{kT}    \right],
\end{equation}
where $\lambda(z) =\delta F_\text{ex}/\delta \rho(z) +v(z)$.
However, the segment density profiles are evaluated
taking into account the constraint imposed by equation~(\ref{eq:con}) of the present work.
We have 
\begin{align}
\rho _{\text{s}j}^{(I)}(z)&=
R_{\text{c}I} \exp [-
\lambda_{j}^{(\text{c}I)}(z)/kT]G_{j}^{(\text{L}I)}(z)G_{M+1-j}^{(\text{R}I)}(z)
 \nonumber\\
&\times\Bigg\{\int_{-H/2}^{H/2} \rd z \exp [-
\lambda_{j}^{(\text{c}I)}(z)/kT]G_{j}^{(\text{L}I)}(z)G_{M+1-j}^{(\text{R}I)}(z)\Bigg\}^{-1},
\label{eq:prof2}
\end{align}
where 
$\lambda _j^{(\text{c}I)}(z)=\delta F_\text{ex}/\delta \rho_\text{s}^{(I)}(z)
+v_{\text{s}j}^{(I)}(z)$ and
where the functions $G_{j}^{(PI)}(z)$, $P=\text{L,\,R}$ are determined from 
the recurrence relations \cite{c21},
\begin{equation}
G_{j}^{(\text{L}I)}(z)=\int \rd z^{\prime }\exp [-\lambda _{j-1}^{(\text{c}I)}(z)/kT]
\frac{\theta (\sigma_\text{c}-|z-z^{\prime }|)}{2\sigma_\text{c}}
G_{j-1}^{(\text{L}I)}(z^{\prime }),
\end{equation}
and
\begin{equation}
G_{j}^{(\text{R}I)}(z)=\int \rd z^{\prime }\exp [-\lambda _{M-j+2}^{(\text{c}I)}(z)/kT]
\frac{\theta (\sigma_\text{c}-|z-z^{\prime }|)}{2\sigma_\text{c}}
G_{j-1}^{(\text{R}I)}(z^{\prime }),
\end{equation}
for $i=2,3,\dots ,M$ and with $G_{1}^{(\text{L}I)}(z)=G_{1}^{(\text{R}I)}(z) \equiv 1$.


\ukrainianpart

\title{Опис адсорбції води у щілиноподібних наноканалах з прищепленими молекулярними щітками. Теорія функціоналу густини}

\author{В.М. Трехос\refaddr{label1}, M. Агілар\refaddr{label2}, С. Соколовські\refaddr{label3}, O. Пізіо\refaddr{label4}}
\addresses{
	\addr{label1} 
	Інститут фундаментальних наук та інженерії, Автономний університет штату Гідальго, Гідальго, Мексика
	\addr{label2}
	Інститут хімії, Національний автономний університет м. Мехіко, Мехіко, Мексика
	\addr{label3} 
	Відділ моделювання фізико-хімічних процесів, університет Марії Склодовської-Кюрі, \\ Люблін 20-031, Польща
	\addr{label4}
	Інститут хімії, Національний автономний університет м. Мехіко, Мехіко, Мексика }

\makeukrtitle 

\begin{abstract}
	Ми дослідили модель адсорбції води у щілиноподібні наноканали з звома стінками, хімічно модифікованими за допомогою прищеплених полімерних шарів, що формують щітки. 	
	В якості теоретичних інструментів використано один з варіантів методу функціоналу густини.
	Модель водоподібного плину, запозичена з роботи Кларка та ін.
	[Mol. Phys., 2006, {\bf 104}, 3561] адекватно відтворює  співіснування пара-рідина в об'ємі. Полімерний шар складається з ланцюжкових молекул в рамках моделі намиста  (pearl-necklace model). Кожна ланцюжкова молекула хімічно зв'язана зі стінками пор одним завершальним сегментом. Основною метою даного дослідження є вивчення залежності висоти полімерного шару від густини прищеплення та мікроскопічної структури  інтерфейсу між
	адсорбованим плином і щітками.  Детально досліджено термічний відгук цих властивостей на адсорбцію. Отримані результати є важливими для розуміння стягування і набрякання молекулярних щіток у наноканалах.
	
	\keywords асоційовані плини, теорія функціоналу густини, адсорбція, молекулярні щітки, модель води
	
\end{abstract}
\lastpage

\begin{thebibliography}{99}

\bibitem{ihor1} Omelyan I.P., Mryglod I.M., Tokarchuk M.V., Condens. Matter Phys., 2005, \textbf{8}, 25, 
\doi{10.5488/CMP.8.1.25}.

\bibitem{ihor2}  Patsahan O.,  Mryglod I., Condens. Matter Phys., 2012, \textbf{15}, 24001, 
\doi{10.5488/CMP.15.24001}.

\bibitem{ihor3} Mryglod I.M., Condens. Matter Phys., 1998, \textbf{1}, 753,  
\doi{10.5488/CMP.1.4.753}.

\bibitem{jonas1} Jonas A.M., Hu Z., Glinel K., Huck W.T.S., Nano Lett., 2008, \textbf{8}, 3819,
\doi{10.1021/nl802152q}.

\bibitem{jonas2}
Jonas A.M., Hu Z., Glinel K., Huck W.T.S., 
Macromolecules, 2008, \textbf{41}, 6859, \doi{10.1021/ma801584k}.

\bibitem{conrad} Conrad J.C., Robertson M.L., Curr. Opin. Solid State Mater. Sci., 
2019, \textbf{23}, 1,
\doi{10.1016/j.cossms.2018.09.004}.

\bibitem{li} Li B., Yu B., Ye Q.,  Zhou F., Acc. Chem. Res., 2015, \textbf{48}, 229,
\doi{10.1021/ar500323p}.

\bibitem{minko1} Minko S., J. Macromol. Sci., Polym. Rev., 
2006, \textbf{46}, 397, \doi{10.1080/15583720600945402}.

\bibitem{minko2}
Uhlmann P., Ionov L., Houbenov N., Nitschke M., Grundke K., Motornov M., Minko S., Stamm M.,
Prog. Org. Coat.,  2006, \textbf{55},  168,
\doi{10.1016/j.porgcoat.2005.09.014}.

\bibitem{minko3}  Brittain W.J.,  Minko S., J. Polym. Sci., Part A: Polym. Chem., 2007, \textbf{45}, 3505, \doi{10.1002/pola.22180}.

\bibitem{slavko1}
Kuroki H., Gruzd A., Tokarev I., Patsahan T.,  Ilnytskyi J., Hinrichs K., Minko S.,
ACS Appl. Mater. Interfaces, 2019, \textbf{11}, 18268,
\doi{10.1021/acsami.9b06679}.

\bibitem{constable1} Constable A.N., Brittain W.J.,  Colloids Surf., A,
2007, \textbf{308},  123,
\doi{10.1016/j.colsurfa.2007.05.059}.

\bibitem{constable2} Constable A.N., Brittain W.J.,  Colloids Surf., A,
2011, \textbf{380},  128,
\doi{10.1016/j.colsurfa.2011.02.030}.

\bibitem{ethier}  Ethier J.G.,  Hall L.M., Macromolecules, 2018, \textbf{51}, 9878,
\doi{10.1021/acs.macromol.8b01373}.

\bibitem{malfreyt} Malfreyt P., Tildesley D.J., Langmuir, 2000, \textbf{16}, 4732,
\doi{10.1021/la991396z}.

\bibitem{dimitrov1} Dimitrov D., Milchev A., Binder K., J. Chem. Phys., 2007, \textbf{127}, 084905,
\doi{10.1063/1.2768525}.

\bibitem{dimitrov2}  Dimitrov D., Milchev A., Binder K., Macromol. Symp., 2007, \textbf{252}, 47,
\doi{10.1002/masy.200750605}.

\bibitem{dimitrov3}  Dimitrov D., Milchev A., Binder K., Heermann D.W., Macromol. Theory Simul.,
2006, \textbf{15}, 573,\\
\doi{10.1002/mats.200600029}.

\bibitem{goicochea}  Goicochea A.G.,  Alarc\'on F., J. Chem. Phys., 2011, \textbf{134}, 014703,
\doi{10.1063/1.3517869}.

\bibitem{ii10}  Grest G.S.,  Murat M., Macromolecules,  1993, \textbf{26}, 3108,
\doi{10.1021/ma00064a019}.

\bibitem{ii11}  Szleifer I.,  Carignano M.A.,  In: Advances in Chemical Physics: Polymeric Systems, Vol. 94,  
Prigogine I., Rice~S.A.~(Eds.), Wiley \& Sons, 2007, 165--260.   

\bibitem{slavko2} Ilnytskyi J., Sokolowski S., Patsahan T., Condens. Matter Phys., 
2013,  \textbf{16}, 13606, \doi{10.5488/CMP.16.13606}.

\bibitem{slavko3}
Ilnytskyi J.M., Patsahan T., Soko{\l}owski S.,  J. Chem. Phys.,  2011, \textbf{134}, 204903,
\doi{10.1063/1.3592562}.

\bibitem{slavko4}
Soko{\l}owski S.,  Ilnytskyi J., Pizio O., Condens. Matter Phys., 2014, \textbf{17}, 12601,
\doi{10.5488/CMP.17.12601}.

\bibitem{milner} Milner S.T., Witten T.A., Cates M.E., Europhys. Lett., 1988, \textbf{5}, 413, \doi{10.1209/0295-5075/5/5/006}.

\bibitem{int6}  Zhulina E.B.,  Singh C.,  Balazs A.C., Macromolecules, 1996, \textbf{29}, 6338,
\doi{10.1021/ma960498i}.

\bibitem{int7}  Zhulina E.B., Leermakers F.,  Borisov O.V., Sci. Tech. J. Inf. Technol. Mech. Opt., 2015, \textbf{15}, 493,
\doi{10.17586/2226-1494-2015-15-3-493-499}.

\bibitem{int8} Zhulina E.B., Leermaker F.A.M., Borisov O.V., Langmuir, 2015, \textbf{31}, 6514,
\doi{10.1021/acs.langmuir.5b00947}.

\bibitem{int9}  Lebedeva I.O.,  Zhulina E.B.,  Borisov O.V., J. Chem. Phys., 2017, \textbf{146}, 214901,
\doi{10.1063/1.4984101}.

\bibitem{chapter} Patrykiejew A., Sokolowski S., Pizio O., In: Surface and Interface Science, Vol. 6,  Solid-Gas Interfaces II, 
Wandelt~K.~(Ed.), Wiley, Berlin, 2016,  883--1253.   

\bibitem{Chapman2007} Jain S.,  Dominik A.,  Chapman W.G., J. Chem. Phys., 2007, \textbf{127}, 244904,
\doi{10.1063/1.2806932}.

\bibitem{Chapman2008} Jain S., Jog P.,  Weinhold J.,  Srivastava R.,  Chapman W.G., J. Chem. Phys., 2008, \textbf{128}, 154910,
\doi{10.1063/1.2902976}.

\bibitem{Chapman2011}  Gong K.,  Chapman W.G., J. Chem. Phys., 2011, \textbf{135}, 214901,
\doi{10.1063/1.3657830}.

\bibitem{Chapman2012}  Gong K.,  Marshall B.D.,  Chapman W.G., J. Chem. Phys., 2012, \textbf{137}, 154904, 
\doi{10.1063/1.4757860}.

\bibitem{Chapman2013}  Gong K.,  Marshall B.D.,  Chapman W.G., J. Chem. Phys., 2013, \textbf{139}, 094904,
\doi{10.1063/1.4819957}.

\bibitem{c40}   Yu Y.-X., Wu J., J. Chem. Phys., 2002, \textbf{117}, 2368, 
\doi{10.1063/1.1491240}.

\bibitem{c41} Yu Y.-X. , Wu  J., J. Chem. Phys., 2002, \textbf{117},  10156,
\doi{10.1063/1.1520530}.


\bibitem{trejos-x1}   Trejos V.M., Pizio O., Sokolowski S., Fluid Phase Equilib., 2018, \textbf{473}, 145,
\doi{10.1016/j.fluid.2018.06.005}.

\bibitem{trejos-x2} Trejos V.M., Soko{\l}owski S., Pizio O., J. Chem. Phys., 2018, \textbf{149}, 134701,
\doi{10.1063/1.5047018}.

\bibitem{trejos-x3} Trejos V.M., Pizio O., Soko{\l}owski S., J. Chem. Phys., 2018, \textbf{149}, 234703,
\doi{10.1063/1.5066552}.

\bibitem{trejos-x4}  Trejos V.M.,  Soko{\l}owski S., Pizio O., Mol. Phys., 2018, \textbf{118}, 1615647,
\doi{10.1080/00268976.2019.1615647}.

\bibitem{trejos-x5}  Trejos V.M., Pizio O., Soko{\l}owski S., J. Chem. Phys., 2019,  \textbf{151}, 064704,
\doi{10.1063/1.5116128}.

\bibitem{clark}  Clark G.N.I.,  Haslam A.J.,  Galindo A.,  Jackson G., Mol. Phys., 2006, \textbf{104}, 3561,\\
\doi{10.1080/00268970601081475}.

\bibitem{gil}  Gil-Villegas A., Galindo A.,  Whitehead P.J.,   Mills S.J., Jackson G.,  Burgess A.N., 
J. Chem. Phys., 1997, \textbf{106}, 4168, 
\doi{10.1063/1.473101}.

\bibitem{Ulberg1994} Ulberg D.E., Gubbins K.E., Mol. Simul., 1994, \textbf{13}, 205,
\doi{10.1080/08927029408021984}.

\bibitem{Ulberg1995} Ulberg D.E., Gubbins K.E., Mol. Phys., 1995, \textbf{84}, 1139,
\doi{10.1080/00268979500100801}.

\bibitem{Cummings2011} Srivastava R., Docherty H., Singh J.K., Cummings P.T., J. Phys. Chem. C,
2011, \textbf{115}, 12448,\\
\doi{10.1021/jp2003563}.

\bibitem{nezbeda1} Kolafa J., Nezbeda I., Mol. Phys., 1987, \textbf{61}, 161,
\doi{10.1080/00268978700101051}.

\bibitem{jackson1}  Jackson G.,  Chapman W.G.,  Gubbins K.E., Mol. Phys., 1988, \textbf{65}, 1,
\doi{10.1080/00268978800100821}.

\bibitem{jackson2}  Chapman W.G.,  Jackson G.,  Gubbins K.E., Mol. Phys., 1988, \textbf{65}, 1057,
\doi{10.1080/00268978800101601}.


\bibitem{steele}   Steele W.A., The Interaction of Gases with Solid Surfaces, Pergamon Press, Oxford, 1974. 

\bibitem{steele2}  Steele W.A., Surf. Sci., 1973, \textbf{36}, 317, \doi{10.1016/0039-6028(73)90264-1}.

\bibitem{caowu}  Cao D.,  Wu J., Langmuir, 2006, \textbf{22}, 2712,
\doi{10.1021/la0527588}.


\bibitem{c21}  Pizio O.,  Bor{\'o}wko M., R{\.z}ysko W., Staszewski T., Soko{\l}owski S., J. Chem. Phys., 2008, \textbf{128},  044702, 
\doi{10.1063/1.2829247}.

\bibitem{c22}   Pizio O.,  Soko{\l}owski S.,  Soko{\l}owska Z.,  J. Chem. Phys., 2011, \textbf{134},  214702,
\doi{10.1063/1.3597773}.

\bibitem{YuWu2002}   Yu Y.-X.,  Wu J., J. Chem. Phys.,  2002, \textbf{116}, 7094,
\doi{10.1063/1.1463435}.


\bibitem{ca2}  Roth R.,  Evans R.,  Lang A.,   Kahl G.,  J. Phys.: Condens. Matter, 2002, \textbf{14}, 12063,\\
\doi{10.1088/0953-8984/14/46/313}.


\bibitem{ca3}  Bryk P.,   Soko{\l}owski S.,  Pizio O., J. Chem. Phys., 2006, \textbf{125}, 024909,
\doi{10.1063/1.2212944}.

\bibitem{wertheim1} Wertheim M.S., J. Stat. Phys., 1984, \textbf{35}, 19, \doi{10.1007/BF01017362}. 
\bibitem{wertheim1_} Wertheim M.S., J. Stat. Phys., 1984, \textbf{35}, 35, \doi{10.1007/BF01017363}.  

\bibitem{wertheim2} Wertheim M.S., J. Stat. Phys., 1986, \textbf{42}, 459, \doi{10.1007/BF01127721}.
\bibitem{wertheim2_} Wertheim M.S., J. Stat. Phys., 1986, \textbf{42}, 477, \doi{10.1007/BF01127722}.  

\bibitem{h1}  Ahrens H.,  F{\"o}rster S.,  Helm C.A., Kumar N.A.,  Naji A.,  Netz R.R.,  Seidel C., 
J. Phys. Chem. B, 2004, \textbf{108}, 16870, 
\doi{10.1021/jp049553c}.
 
\bibitem{orest} Pizio O., In: Computational Methods in Surface and Colloid Science, Bor\'owko M.~(Ed.), Marcel Dekker, New York, 2000, 293--346.  

\bibitem{jps} Pizio O., Sokolowski S., J. Phys. Stud., 1998, \textbf{2}, 296. 

\bibitem{stah} Bor\'owko M., R\.zysko W., Soko{\l}owski S., Staszewski T., 
J. Phys. Chem. B,  2009,  \textbf{113},  4763,\\
\doi{10.1021/jp811143n}.

\bibitem{kasia}
Bucior K., Patrykiejew A., Pizio O., Soko{\l}owski S., J. Colloid Interface Sci.,  2003, \textbf{259}, 209,\\
\doi{10.1016/S0021-9797(02)00203-5}.

\end{thebibliography}
\end{document}